\documentclass[usegraphicx,usenatbib,useAMS]{mn2e}

\usepackage{amssymb}
\usepackage{soul}

\newcommand{\aap}{A\&A}

\newcommand{\aaps}{A\&AS}
\newcommand{\aj}{AJ}
\newcommand{\apj}{ApJ}
\newcommand{\apjs}{ApJS}
\newcommand{\apss}{Ap\&SS}
\newcommand{\aspc}{ASP Conf. Ser.}
\newcommand{\mnras}{MNRAS}
\newcommand{\na}{New Astronomy}%%%%%%%%%%%%%
\newcommand{\pasj}{PASJ}
\newcommand{\pasp}{PASP}
\newcommand{\jcp}{J. Chem. Phys.}
\newcommand{\jms}{J. Mol. Spectrosc.}
\newcommand{\jmst}{J. Mol. Struct.}

\newcommand{\pccp}{Phys. Chem. Chem. Phys.}

\newcommand{\sa}{Spectrochimica Acta}
\newcommand{\nature}{Nature}

\newcommand{\subscript}[1]{\textnormal{\scriptsize{#1}}}
\newcommand{\rstar}{\ensuremath{R_\star}}
\newcommand{\rin}{\ensuremath{R_\subscript{in}}}
\newcommand{\rout}{\ensuremath{R_\subscript{out}}}

\newcommand{\kms}{km~s$^{-1}$}

\newcommand{\klam}{k$\lambda$}
\newcommand{\phasecalib}{0854+201}
\newcommand{\bandpasscalib}{3C84}
\newcommand{\irc}{IRC+10216}

\usepackage{xcolor}

\setul{0pt}{1pt}
\begin{document}

\title[Molecules in the dust formation zone of \irc]
{The complex dust formation zone of the AGB star \irc{} probed with
CARMA 0.25~arcsec angular resolution molecular observations}
\author[J. P. Fonfr\'ia et al.]{J. P. Fonfr\'ia,$^1$\thanks{E-mail: fonfria@astro.unam.mx}
M. Fern\'andez-L\'opez,$^{2,3}$ M. Ag\'undez,$^4$ C. S\'anchez-Contreras,$^5$
\newauthor
S. Curiel$^6$ and J. Cernicharo$^4$\\
$^1$ Departamento de Estrellas y Medio Interestelar, Instituto de Astronom\'ia, 
Universidad Nacional Aut\'onoma de M\'exico, 
Ciudad \\
Universitaria, 04510, Mexico City (Mexico)\\
$^2$ Instituto Argentino de Radioastronom\'ia, CCT-La Plata (CONICET), C.C.5,
1894, Villa Elisa (Argentina)\\
$^3$ Department of Astronomy, University of Illinois at Urbana-Champaign, 
1002 West Green Street, Urbana, IL 61801 (USA)\\
$^4$ Laboratorio de Astrof\'isica Molecular, Departamento de Astrof\'isica,
Centro de Astrobiolog\'ia, INTA-CSIC, 28850
Torrej\'on de Ardoz, \\
Madrid (Spain)\\
$^5$ Department of Astrophysics, Astrobiology Center (CSIC-INTA), 
Postal address: ESAC campus, P.O. Box 78, E-28691, Villanueva\\
de la Ca\~nada, Madrid (Spain)\\
$^6$ Departamento de Astrof\'isica Te\'orica, Instituto de Astronom\'ia, 
Universidad Nacional Aut\'onoma de M\'exico, 
Ciudad Universitaria, \\
04510, Mexico City (Mexico)}

\maketitle

\begin{abstract}
We present low spectral resolution
molecular interferometric observations at 1.2~mm obtained with 
the Combined Array for Research in Millimetre-wave Astronomy (CARMA) towards 
the C-rich AGB star \irc.
We have mapped the emission of several lines of SiS, H$^{13}$CN, SiO, and 
SiC$_2$ in the ground and first excited vibrational states with a
high angular resolution of 0.25~arcsec.
These observations have allowed us to partially resolve the emission of the 
envelope at distances from the star $\lesssim 50$~stellar radii (\rstar),
where the stellar wind is mainly accelerated.
The structure of the molecular emission
has been modelled with a 3D radiation transfer code.
The emission of line SiS($v=0,J=14-13$) is best reproduced 
with a set of maser emitting arcs arranged between 5 and 20~\rstar.
The abundance of H$^{13}$CN with respect to H$_2$ 
decreases from $8\times 10^{-7}$ 
at $1-5$~\rstar{} to $3\times 10^{-7}$ at 20~\rstar.
The SiO observations are explained with an abundance 
$\lesssim 2\times 10^{-8}$ in the shell-like region
between 1 and 5~\rstar.
At this point, the SiO abundance sharply increases up to
$(2-3)\times 10^{-7}$.
The vibrational temperature of SiO increases by a factor of 2 due 
North-East between 20 and 50~\rstar.
SiC$_2$ is formed at the stellar surface with an abundance of 
$8\times 10^{-7}$
decreasing down to 
$8\times 10^{-8}$ at 20~\rstar{}
probably due to depletion on to dust grains.
Several asymmetries are
found in the abundance distributions of H$^{13}$CN, SiO, and
SiC$_2$ which define three remarkable directions
(North-East, South-Southwest, and South-East) in the explored 
region of the envelope.
There are some differences between the red- and blue-shifted emissions of
these molecules suggesting the existence of additional asymmetries in their 
abundance distributions along the line-of-sight.
\end{abstract}
\begin{keywords}
stars: AGB and post-AGB --
stars: individual (\irc) --
stars: abundances --
circumstellar matter --
masers --
techniques: interferometric
\end{keywords}

\section{Introduction}

The asymptotic giant branch star (AGB) \irc, considered the archetypical star 
of this type, has been extensively observed since its discovery 
\citep{becklin_1969}.
Its C-rich circumstellar envelope (CSE) is mainly composed of dust and 
molecular gas.
A large amount of molecules have been observed to date in this envelope
\citep*[e.g.,][]{kawaguchi_1995,cernicharo_2000,he_2008}.
Among these molecules, just a few are formed close to the stellar 
photosphere \citep*[the so-called parent molecules such as C$_2$H$_2$, HCN, 
SiS, SiO, SiC$_2$, and CS; e.g.][]{keady_1993},
while the rest arise in the outer shells, where the Galactic UV radiation is 
able to trigger a very active chemistry due to the small dust opacity.

The great sensitivity of the millimetre telescopes built in the last decades 
has been crucial to understand the chemistry in the outer envelope of \irc.
However, in spite of its proximity 
\citep[$123\pm 14$~pc;][]{groenewegen_2012},
the small solid angle subtended by the innermost shells of its CSE has 
prevented from achieving a deeper comprehension of the chemical evolution of 
the gas near the star, where even O-bearing molecules such as H$_2$O, typical 
in O-rich envelopes and previously believed to be marginally produced in 
C-rich ones, are observed with abundances larger than expected
\citep{willacy_1998,melnick_2001,agundez_2006,cherchneff_2006,agundez_2010,
decin_2010b,neufeld_2011}.

Previous works based on millimetre interferometric observations
\citep{bieging_1993,lucas_1995,monnier_2000b,
young_2004,schoier_2006b,schoier_2007,patel_2009,shinnaga_2009},
millimetre and submillimetre single-dish observations 
\citep{turner_1987,fonfria_2006,schoier_2006a,cernicharo_2010,cernicharo_2011,
decin_2010a,decin_2010b,agundez_2012}, and infrared observations
\citep{keady_1988,keady_1993,boyle_1994,schoier_2006b,
fonfria_2008} suggest that the abundance distributions of most
of the parent molecules display essentially spherical symmetry.
None the less, near infrared continuum observations carried out with 
different techniques and high angular resolution 
\citep*[up to $\sim 1$~mas;][]{ridgway_1988,kastner_1994,sloan_1995,
haniff_1998,weigelt_1998,richichi_2003,leao_2006,menut_2007}, 
indicate that the innermost dusty envelope displays a global bipolar morphology
with its major axis roughly aligned along the NE-SW direction 
(P.A.~$\simeq 20\degr$) and a size $\simeq 0.5\times 0.3$~arcsec$^2$
\citep{menshchikov_2001,menshchikov_2002}.
The dusty envelope is also extremely clumpy and undergoes rapid time variations 
\citep{monnier_1998,osterbart_2000,tuthill_2000,weigelt_2002,males_2012}.
Hence, a complex chemical evolution of the gas linked to the dynamics 
of the inner layers of the CSE is expected 
\citep{willacy_1998,agundez_2006,cherchneff_2006}.

In this paper, we report new molecular observations at 1.2~mm towards \irc{} 
with angular resolutions as high as $0.25$~arcsec.
The observations were performed with the Combined Array for Research in 
Millimeter-wave Astronomy (CARMA)\footnote{Support for CARMA construction was 
derived from the states of Illinois, California, and Maryland, the James S. 
McDonnell Foundation, the Gordon and Betty Moore Foundation, the Kenneth T. and 
Eileen L. Norris Foundation, the University of Chicago, the Associates of 
the California Institute of Technology, and the National Science Foundation. 
Ongoing CARMA development and operations are supported by the National 
Science Foundation under a cooperative agreement, and by the CARMA partner 
universities. } and include several lines of SiS, H$^{13}$CN, SiO, and SiC$_2$.
In Section~\ref{sec:observations} we introduce the observations. 
The observational results derived from the analysis 
of the continuum and molecular emission are shown in 
Section~\ref{sec:observational.results}. 
The observed molecular brightness distributions are fitted in 
Section~\ref{sec:modelling}.
Section~\ref{sec:discussion} contains the analysis of the fits, the comparison 
of the results with those of previous works, and a discussion of their 
implication.
Finally, we present our main conclusions in Section~\ref{sec:conclusions}.

\subsection{The envelope of \irc{} and its molecular content}
\label{sec:previous}

\irc{} is composed of a central variable star with a pulsation period 
$\simeq 625-650$~days 
\citep{witteborn_1980,ridgway_1988,jones_1990,dyck_1991,lebertre_1992,
jenness_2002} and an angular radius, $\alpha_\star$, 
that ranges roughly between 15 and 
25~mas along the pulsation \citep{ridgway_1988,monnier_2000a,menshchikov_2001}.
Its LSR systemic velocity has been accurately estimated in $-26.5$~\kms{} 
\citep*[e.g.][]{guelin_1993,loup_1993,he_2008,cernicharo_2011}.
It is commonly accepted that the massive condensation of the refractory 
molecular species on to dust grains occurs in the so-called dust formation zone,
which extends roughly from 5 up to about $20-50~\rstar$, where the 
terminal expansion velocity is reached 
\citep*[$\simeq 14.5$~\kms; e.g.,][]{cernicharo_2000,he_2008}.
The expansion velocity field in this region of the CSE is not accurately
determined and two different scenarios are usually adopted:
($i$) the gas is continuously accelerated from an expansion velocity of
$2-3$~\kms{} at $\simeq 5~\rstar$,
reaching the terminal velocity at $40-50~\rstar$ 
\citep*[e.g.,][]{schoier_2006b,decin_2010a}, or
($ii$) the gas is mostly accelerated in two $\sim 1~\rstar$ width shells
located at $\simeq 5$ and $20~\rstar$
(inner and outer acceleration shells/zones, hereafter),
where the expansion velocity is increased from 5 to 11~\kms{} and
from 11~\kms{} to the terminal velocity, respectively.
The line width is usually assumed to be
$\simeq 5$~\kms{} at the stellar surface due to turbulence, 
decreases down to $\simeq 1$~\kms{} 
around the inner acceleration zone, and remains constant 
outwards \citep{keady_1988,keady_1993,loup_1993,boyle_1994,fonfria_2008,
cernicharo_2011,agundez_2012}.
Other scenarios are also possible, such as an irregular 
gas expansion velocity field linked to dust clumps moving close to the star 
\citep*[e.g.,][]{menut_2007}.

SiS is one of the most abundant species in the innermost envelope of \irc{}
\citep*[][Fonfr\'ia et al., in preparation]{keady_1993,boyle_1994,agundez_2012}.
The drop observed in its abundance profile across the dust formation zone is 
usually attributed to the refractory nature of this Si-bearing molecule, which 
would favour its deposition on to the dust grains 
\citep{boyle_1994,decin_2010a,agundez_2012}.
At scales larger than 5~arcsec the brightness distribution is roughly spherical 
\citep{bieging_1989,bieging_1993}, while higher angular resolution observations
(HPBW~$\simeq 3$~arcsec) reveal an elongation along the direction with a
P.A.~$\simeq 20\degr$ \citep{lucas_1995}.
Maser emission has been detected in several rotational lines coming from 
the dust formation zone \citep{henkel_1983,carlstrom_1990,fonfria_2006}.
One of them, SiS($v=0,J=14-13$), is analysed in the current work.

H$^{13}$CN is probably formed at the stellar photosphere, where emission
of very high excitation ro-vibrational levels of HCN have been observed 
\citep{cernicharo_2011} and thermodynamical equilibrium (TE) is supposed to 
prevail \citep*[e.g.,][]{willacy_1998,agundez_2006}.
The abundance profile derived from single-dish and interferometric
observations displays a somewhat complex dependence on the distance to the 
star before $\simeq 20-50$~\rstar{} from where the abundance remains 
nearly constant 
\citep{schoier_2007,fonfria_2008,shinnaga_2009}.

\begin{table*}
\begin{minipage}{\textwidth}
\caption{Summary of observations at 257~GHz towards \irc}
\label{tab:table1}
\begin{tabular}{c@{\hspace{5ex}}c@{\hspace{5ex}}c@{\hspace{5ex}}c@{\hspace{5ex}}c@{\hspace{5ex}}cccc}
\hline
Epoch & Configuration & Baselines & Time$^a$ & $\tau_{230}$ & Phase Calibrator Flux$^b$ & \multicolumn{2}{c}{\hrulefill Synthesised Beam$^c$\hrulefill} & RMS$^c$ \\
      &         &           &          &             &                    & HPBW                     & P.A.         & \\
      &         & (m)       & (hrs)    &             & (Jy)               & (arcsec$^2$) & ($\degr$)    & (mJy~beam$^{-1}$)\\
\hline
2011 Jan 5 & B & $60-800$ & 3.1 & 0.07 & 3.8 & $0.3\times0.2$ & 57 & 10 \\
2012 Mar 8 & C & $20-300$ &  3.9 & 0.12 & 3.4 & $0.7\times0.6$ & 89  & 10 \\
\hline
\end{tabular}
\medskip
\newline
($a$) On source integration time after removal of bad weather periods.
($b$) Flux set to the phase calibrator \phasecalib. It was estimated from the 
CARMA and the SMA quasar monitoring.
($c$) Estimated from the continuum images.
\end{minipage}
\end{table*}

\begin{figure*}
\includegraphics[width=\textwidth]{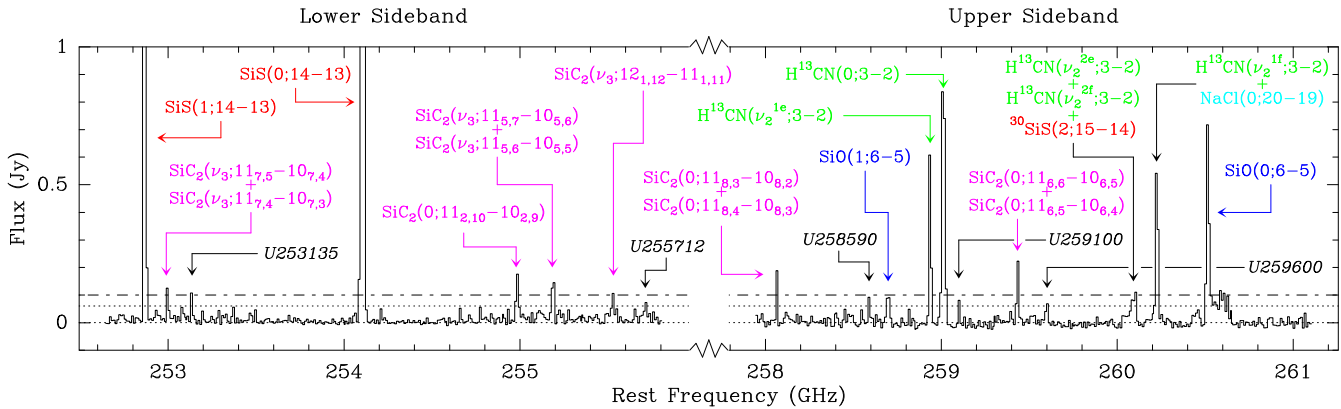}
\caption{Observed spectrum of \irc{} in B configuration.
It contains lines of SiS (red), H$^{13}$CN (green), SiO (blue), SiC$_2$ 
(magenta), and several unidentified features (black).
The lower dotted line indicate the baseline after the continuum removal, the 
upper one indicates the $3\sigma$ level, and the dash-dotted line corresponds 
to the $5\sigma$ level.
The rms of the noise ($\sigma\simeq 20$~mJy) has been estimated from frequency 
ranges where the spectrum is apparently free of lines.
The molecular transitions associated to the observed lines are labelled with the 
vibrational state in which they are produced (within the parentheses, to the 
left) and the rotational quantum numbers of lower and upper levels, 
$J_\subscript{up}-J_\subscript{low}$ (within the parentheses, to the right).
The overlaps between lines have been designated by means of the symbol $+$.
The unidentified lines have been denoted in the usual way.
}
\label{fig:f1}
\end{figure*}

SiO seems to be formed near the star as well \citep{willacy_1998,agundez_2006}.
The abundance distribution in the inner shells of the envelope is still a 
controversial topic since some works derived constant abundances 
\citep{keady_1993,schoier_2006a,agundez_2012}
while others found more complex distributions in which there is a significant 
increase of the abundance at distances to the star $\lesssim 8$~\rstar{}
\citep{schoier_2006b,decin_2010a}.
All the observations carried out to date (HPBW~$\gtrsim 2$~arcsec) are 
compatible with a roughly spherical abundance distribution.

The emission of SiC$_2$ is composed of a bright compact region surrounding 
the central star and a detached and clumpy shell with a radius of about 
15~arcsec, which is apparently hollow and displays a marked bipolar emission
along the NE-SW direction 
\citep{takano_1992,gensheimer_1995,lucas_1995,cernicharo_2010}.
This brightness distribution suggests that SiC$_2$ is formed near the star, 
gets depleted in the dust formation zone, and again reappears in the outer 
CSE probably due to the action of the UV Galactic radiation field on the 
expanding matter.

\section{Observations}
\label{sec:observations}

\begin{table*}
\begin{minipage}{\textwidth}
\caption{Parameters of observed lines in B configuration}
\label{tab:table2}
\begin{tabular}{c@{\hspace{4.7ex}}cr@{$-$}lc@{\hspace{4.7ex}}c@{\hspace{4.7ex}}c@{\hspace{4.7ex}}c@{\hspace{4.7ex}}c@{\hspace{4.7ex}}c}
\hline
Molecule & \multicolumn{3}{c}{\hrulefill\hspace{1ex}Transition\hspace{1ex}\hrulefill} & $\nu_\textnormal{\tiny rest}$ & $E_\subscript{up}/k$ & $V_\textnormal{\tiny LSR}$ & FWHM   & Peak flux & Integrated flux \\
         & Vib. State    & $J_\subscript{up}$     & $J_\subscript{low}$                   &                             &                    &                          &        &           &  \\
         &               & \multicolumn{2}{c}{}                                       & (MHz)                       & (K)                & (\kms)                   & (\kms) & (Jy)      & (Jy~\kms)\\
\hline
\multicolumn{10}{c}{\textit{Identified lines}}\\
\hline
SiS              & $v=0$        & $14$     & $13$       & $254103.2108\phantom{^a}$    & $91.5$       & $-25\pm 6$ & $18\pm 6$ & $13.12\pm 0.02$ &  $250\pm 80$ \\
                 & $v=1$        & $14$     & $13$       & $252866.4683\phantom{^a}$    & $1162.2$      & $-27\pm 6$ & $17\pm 6$ & $2.25\pm 0.02$ &  $40\pm 15$ \\
$^{30}$SiS       &  $v=2$        & $15$     & $14$       & $260074.7218{}^a$            & $2196.9$     & $-31\pm 6$ & $6\pm 6$ & $0.24\pm 0.02$  &  $1.5\pm 1.7$ \\
H$^{13}$CN       & $v=0$        & $3$       & $2$        & $259011.7978\phantom{^a}$   & $24.9$       & $-26\pm 6$ & $21\pm 6$ & $1.01\pm 0.02$  &  $23\pm 7$ \\
                & $\nu_2^{1e}$  & $3$       & $2$        & $258936.0502\phantom{^a}$   & $1040.6$      & $-27\pm 6$ & $14\pm 6$ & $0.70\pm 0.02$  &  $10\pm 5$ \\
                & $\nu_2^{1f}$    & $3$       & $2$        & $260224.8132\phantom{^a}$   & $1040.7$      & $-26\pm 6$ & $18\pm 6$ & $0.62\pm 0.02$  & $12\pm 4$ \\
                & $2\nu_2^{2e}$  & $3$       & $2$        & $260094.0^*{}^a\phantom{00}$    & $2964^*$        & $-32\pm 6$ & $6\pm 6$ & $0.16\pm 0.02$  &  $1.0\pm 1.2$ \\
                & $2\nu_2^{2f}$    & $3$       & $2$        & $260104.2^*{}^a\phantom{00}$    & $2964^*$        & $-32\pm 6$ & $6\pm 6$ & $0.16\pm 0.02$  &  $1.0\pm 1.2$ \\
SiO             & $v=0$        & $6$       & $5$        & $260518.0200\phantom{^a}$   & $43.8$       & $-24\pm 6$ & $18\pm 6$ & $0.73\pm 0.02$  &  $14\pm 5$ \\
                & $v=1$        & $6$       & $5$        & $258707.3900\phantom{^a}$   & $1812.6$     & $-28\pm 6$ & $26\pm 6$ & $0.10\pm 0.02$  &  $2.8\pm 1.2$ \\
 SiC$_2$        & $v=0$        & $11_{2,10}$ & $10_{2,9}$ & $254981.4940\phantom{^a}$    & $81.9$      & $-29\pm 6$ & $20\pm 6$ & $0.17\pm 0.02$  &  $3.6\pm 1.5$ \\
                & $v=0$        & $11_{8,3}$ & $10_{8,2}$ & $258065.0545^b$              & $199.4$      & $-23\pm 6$ & $9\pm 6$ & $0.26\pm 0.02$  &  $2.5\pm 1.8$ \\
                & $v=0$        & $11_{8,4}$ & $10_{8,3}$ & $258065.0545^b$              & $199.4$      &  &  &  &  \\
                & $v=0$        & $11_{6,6}$ & $10_{6,5}$ & $259433.3090^c$              & $144.9$      & $-27\pm 6$ & $14\pm 6$ & $0.23\pm 0.02$ &  $3.4\pm 1.8$ \\
                & $v=0$        & $11_{6,5}$ & $10_{6,4}$ & $259433.3090^c$              & $144.9$      &  &  &  &   \\
                & $\nu_3$      & $11_{7,5}$ & $10_{7,4}$ & $252998.2722^d$              & $454.8$      & $-27\pm 6$ & $10\pm 6$ & $0.18\pm 0.02$  &  $1.9\pm 1.4$ \\
                & $\nu_3$      & $11_{7,4}$ & $10_{7,3}$ & $252998.2743^d$              & $454.8$      &  &  &   &  \\
                & $\nu_3$      & $11_{5,7}$ & $10_{5,6}$ & $255180.2915^e$              & $406.5$      & $-28\pm 6$ & $17\pm 6$ & $0.16\pm 0.02$ &  $2.9\pm 1.4$ \\
                & $\nu_3$      & $11_{5,6}$ & $10_{5,5}$ & $255191.7263^e$              & $406.5$      &  &  &  &   \\
                & $\nu_3$      & $12_{1,12}$ & $11_{1,11}$ & $255525.3490\phantom{^a}$  & $365.1$      & $-28\pm 6$ & $18\pm 6$ & $0.10\pm 0.02$  & $1.9\pm 1.0$\\
\hline
\multicolumn{10}{c}{\textit{Unidentified lines}}\\
\hline
U253135        &  \multicolumn{3}{c}{} & 253135$\phantom{.0000^a}$  & \multicolumn{2}{c}{}         & $12\pm 6$ & $0.10\pm 0.02$ & $1.3\pm 0.9$ \\
U255712        &  \multicolumn{3}{c}{} & 255712$\phantom{.0000^a}$  & \multicolumn{2}{c}{}          & $28\pm 6$ & $0.06\pm 0.02$ & $1.8\pm 1.0$ \\
U258590        &  \multicolumn{3}{c}{} & 258590$\phantom{.0000^a}$  & \multicolumn{2}{c}{}          & $10\pm 6$ & $0.12\pm 0.02$ & $1.3\pm 1.0$ \\
U259100        &  \multicolumn{3}{c}{} & 259100$\phantom{.0000^a}$  & \multicolumn{2}{c}{}          & $14\pm 6$ & $0.10\pm 0.02$ & $1.5\pm 0.9$ \\
U259600        &  \multicolumn{3}{c}{} & 259600$\phantom{.0000^a}$  & \multicolumn{2}{c}{}          & $18\pm 6$ & $0.08\pm 0.02$ & $1.5\pm 0.9$\\
\hline
\end{tabular}
\medskip
\newline
Most of the rest frequencies and the energies of the upper levels of the 
transitions have been taken from The CDMS Catalogue \citep{muller_2005}.
Those marked with symbol $^*$ have been adopted from \citet{cernicharo_2011}.
We have performed Gaussian fits to the lines with the aid of the GreG package 
included in \textsc{gildas}. 
From these fits, we have retrieved the velocity of the peak flux 
($V_\textnormal{\tiny LSR}$), the line width of the observed lines (FWHM), the 
peak flux (Peak flux), and the area of the observed lines (Integrated flux).
The uncertainties of magnitudes $V_\textnormal{\tiny LSR}$ and FWHM have been
assumed to be the half of a channel ($\simeq 6$~\kms).
We have adopted the rms of the spectrum as the error of the peak emission 
($\simeq 20$~mJy).
The uncertainties of the flux density has been calculated from the latter 
errors.
The lines whose frequencies go with the same letter ($a$, $b$, $c$, \ldots) 
are blended.
The parameters of the blended lines have been estimated during the fitting
procedure by assuming the lines have similar emissions.
\end{minipage}
\end{table*}

The observations analysed in this work were carried out on 2011 January 5 and 
2012 March 8 with the CARMA interferometer (Table~\ref{tab:table1}).
The array was used in its 15-antenna mode.
It comprised six 10.4~m and nine 6.1~m antennas that were arranged in B and 
C configurations, with baselines in the ranges $70-800$ and $20-300$~\klam{} 
and system temperatures of about 200 and 500~K, respectively.
The weather was good during both observing runs, with an atmospheric opacity 
$\simeq 0.1$ at 230~GHz.

The correlator setup included 16 spectral windows (8 windows per sideband 
symmetrically distributed from the first local oscillator frequency), each with 
a configurable band width. 
The frequency of the first local oscillator was tuned at $\simeq 257$~GHz.
In B configuration, the correlator setup included fourteen spectral
windows of 500~MHz and a spectral resolution of 10.4~MHz 
($\simeq 12.0-12.3$~\kms{} at the observed frequency). 
These wide bands covered the line emission from several
molecular species (Figure~\ref{fig:f1}, Table~\ref{tab:table2}). 
In C configuration, four
500~MHz bands were included and the main lines detected in
B configuration were covered with narrower 62~MHz bands (spectral
resolution of 0.2~\kms{} at the observed frequency). 
To combine the data from both configurations we
binned the C configuration data into 12.3~\kms{} channels. 
The high spectral resolution observations will be presented elsewhere.

The calibration and data reduction were performed in the standard way using 
the \textsc{Miriad} package. 
Bandpass solutions were obtained observing \bandpasscalib, while \phasecalib{} 
was periodically observed to calibrate phases. 
Calibrator \phasecalib{} was assumed to have flux densities of 3.8 and 3.4~Jy 
at 257~GHz in 2011 January and 2012 March, respectively, as determined by the 
CARMA and SMA quasar monitoring programs near the time of our observations.
The absolute flux calibration uncertainty is about 15~per~cent.

The phase centre displayed an offset between B and C configuration observations.
In order to combine the data from both configurations
at the inverse Fourier transform step, we shifted the 
coordinates of the phase centre of the data from C configuration using the 
\textsc{Miriad} task \textsc{uvedit}. 
We applied a natural weighting to all data sets and obtained B, C, and
combined BC maps of \irc.
Maps obtained from B configuration data have a synthesised beam of about 
$0.25$~arcsec, while those obtained from C configuration data have a synthesised 
beam of $0.7$~arcsec. 
Combined maps resulted in intermediate resolutions 
(HPBW~$\simeq 0.38-0.46$~arcsec) depending on the \texttt{robust} parameter
\citep{briggs_1995,briggs_1999}.
This parameter weights the data according to the number of visibilities 
measured in each $uv$ space cell. 
In \textsc{Miriad}, the parameter \texttt{robust}~$\gtrsim 2$ and $\lesssim -1$
corresponds to natural and uniform weighting, respectively. 
The natural weighting produces maps with lower rms noise level while the 
uniform weighting provides better angular resolution at the expense of 
increasing sidelobe levels. 
Values between these limits give a compromise between both.
The primary beam of the observations was 30~arcsec for both configurations.
The imaging and analysis of the data were done
using the packages \textsc{Miriad} and \textsc{gildas}.

In this work we present and analyse the line-free continuum data and the
spectral data from five molecular lines observed in both B and C configurations:
SiS($v=0,J=14-13$), H$^{13}$CN($v=0,J=3-2$), SiO($v=0,J=6-5$), SiO($v=1,J=6-5$)
and SiC$_2$($v=0,J_{K_a,K_b}=11_{2,10}-10_{2,9}$).

\subsection{Extended emission}
\label{sec:extended.emission}

It could be argued that the lack of short baselines in our observations would
critically affect the derived maps invalidating our results.
In fact, the absence of these baselines prevents us from detecting emission 
coming from structures with large characteristic lengths which could be 
significant.
If we adopt a Gaussian to roughly represent the smallest structure
missed by C configuration, it would be described by a FWHM in the $uv$ plane 
of $\simeq 50$~\klam.
This Gaussian would have a FWHM in the image plane of about 4~arcsec in 
average (FWHM$_\subscript{image plane}$(arcsec) 
$\simeq 182/$FWHM$_\subscript{$uv$ plane}$(\klam)).
The lack of such structure in our observations compared with other 
observations with a better coverage of the $uv$ plane close to its origin 
would be reflected in
($i$) a smaller emission coming from the surroundings of the star, and
($ii$) a significantly lesser flux density in the regions of the brightness 
distribution at distances from the star $\gtrsim 2$~arcsec.
However, the small features in the emission would not be modified since that 
Gaussian structure would not have a meaningful contribution to the visibility 
at long baselines.
Hence, since we are interested in the shape of the observed distributions
closer to the star than 1~arcsec, the lack of short baselines is not an issue
and we do not expect serious modifications with respect to the actual 
brightness distributions.

\section{Observational results}
\label{sec:observational.results}

\subsection{Continuum emission}
\label{sec:continuum}

The brightness distribution of the continuum of \irc{} at 257~GHz is compatible 
with a point source in B and C configurations.
The integrated flux density is $\simeq 0.38$, and $0.53$~Jy for the maps in B 
and C configurations, respectively, with an uncertainty of 15~per~cent.
Some unresolved observations of the source at different frequencies with 
similar point spread functions (PSFs) to that of our C configuration 
observations can be found in the literature:
\citet{lucas_1999} reported the fluxes $\simeq 0.32$, 0.41, and 0.49~Jy at 216, 
235, and 242~GHz, respectively, with the PdBI and a HPBW~$\simeq 0.5$~arcsec,
\citet{lucas_1997} derived a flux $\simeq 0.48$~Jy at 233~GHz also with the 
PdBI and a HPBW~$\simeq 0.75$~arcsec, and \citet{patel_2009} obtained $0.84$ and 
$1.17$~Jy at 301.1 and 337.5~GHz, respectively, with the SMA and a 
HPBW~$\simeq 0.7$~arcsec.
All these data can be used to estimate the spectral index of the SED of \irc.
The pulsation phase during our observations was $0.5$ \citep{jenness_2002},
while it was $\simeq 0.2$ and 0.3 when the observations by \citet{patel_2009} 
and \citet{lucas_1997} happened.
\citet{lucas_1999} did not indicate the date of their observations.
After correcting the fluxes from the different phase with the aid of the time 
dependent emission formula at $850~\mu$m proposed by \citet{jenness_2002}, we 
derive a spectral index $\simeq 2.53\pm 0.15$.
The uncertainties in the flux calibration rises this error to 0.7.
Thus, this spectral index is compatible with that of a black-body
\citep{young_2004,menten_2006}.

\subsubsection{Source position and proper motions}
\label{sec:source.position}

The position of the continuum source (which we will assume the same as that of 
the central star, hereafter) can be determined after calibration by fitting a 
Gaussian to the flux density distribution.
From the B configuration data we found that the star is at 
$\alpha($J$2000)=09$h$47$m$57.435\textnormal{s}\pm0.002$s and
$\delta($J$2000)=13\degr16'43\farcs86\pm0\farcs03$,
in good agreement with the results previously reported by \citet{patel_2009}.
The uncertainty of this position has been estimated by adding the statistical 
and systematic uncertainties in quadrature 
\citep*[see e.g.,][]{downes_1999,maness_2008}.
The former is
\begin{equation}
\Delta\theta\simeq\left(\frac{4}{\pi}\right)^\frac{1}{4} 
\left(8\ln2\right)^\frac{1}{2} 
\frac{\theta_\subscript{beam}}{\textnormal{SNR}}\quad,
\end{equation}
where $\theta_\subscript{beam}$ is the HPBW of the synthesised beam and SNR is 
the signal-to-noise ratio of the observed source. 
In our case, the statistical error is $0\farcs01$. 
Regarding the systematic positional uncertainty, we have used two independent 
methods to estimate it.
In the first one, we measure the position of the peak emission of \irc{} in each
observing cycle, after the data had been calibrated with \phasecalib{} but 
before self-calibration, to avoid information losses on the position of the 
source \citep*[e.g.][]{menten_2006}. 
In the second method, we determine the rms of the visibility phases of the 
calibrator \phasecalib{} for baselines longer than 200~m, which provide the 
positional accuracy for the most compact sources, such as \irc.
Both methods agree quite well resulting in a positional error due to 
systematics of $0\farcs03$. 
The total positional uncertainty for the continuum source is thus 
$\simeq 0\farcs03$.

This peak position is accurate enough to derive the proper motions of the 
continuum source of \irc{} comparing it 
with the measurements published previously in the literature.
We have used the continuum position and the uncertainties obtained by
\citet{drake_1991} and \citet{menten_2006}, which are based on centimetre
observations (1.3, 2, 3.5, and 6~cm) carried out with the VLA (some of them in 
the A configuration) between years 1987 and 1993.
In order to estimate the proper motions we have assumed that the continuum peak 
emission in the cm and mm wavelength ranges is produced at the same place.
The proper motions of \irc{} derived in the frame of this work are 
$(\mu_\alpha,\mu_\delta)=(34\pm4,12\pm4)$~mas~yr$^{-1}$, in good agreement with 
$(26\pm6,4\pm6)$~mas~yr$^{-1}$ derived by \citet{menten_2006} and, particularly,
with $(35\pm1,12\pm1)$~mas~yr$^{-1}$, recently proposed by \citet{menten_2012}.

\subsubsection{Structure of the continuum source}
\label{sec:continuum.structure}

\begin{figure}
\includegraphics[width=0.475\textwidth]{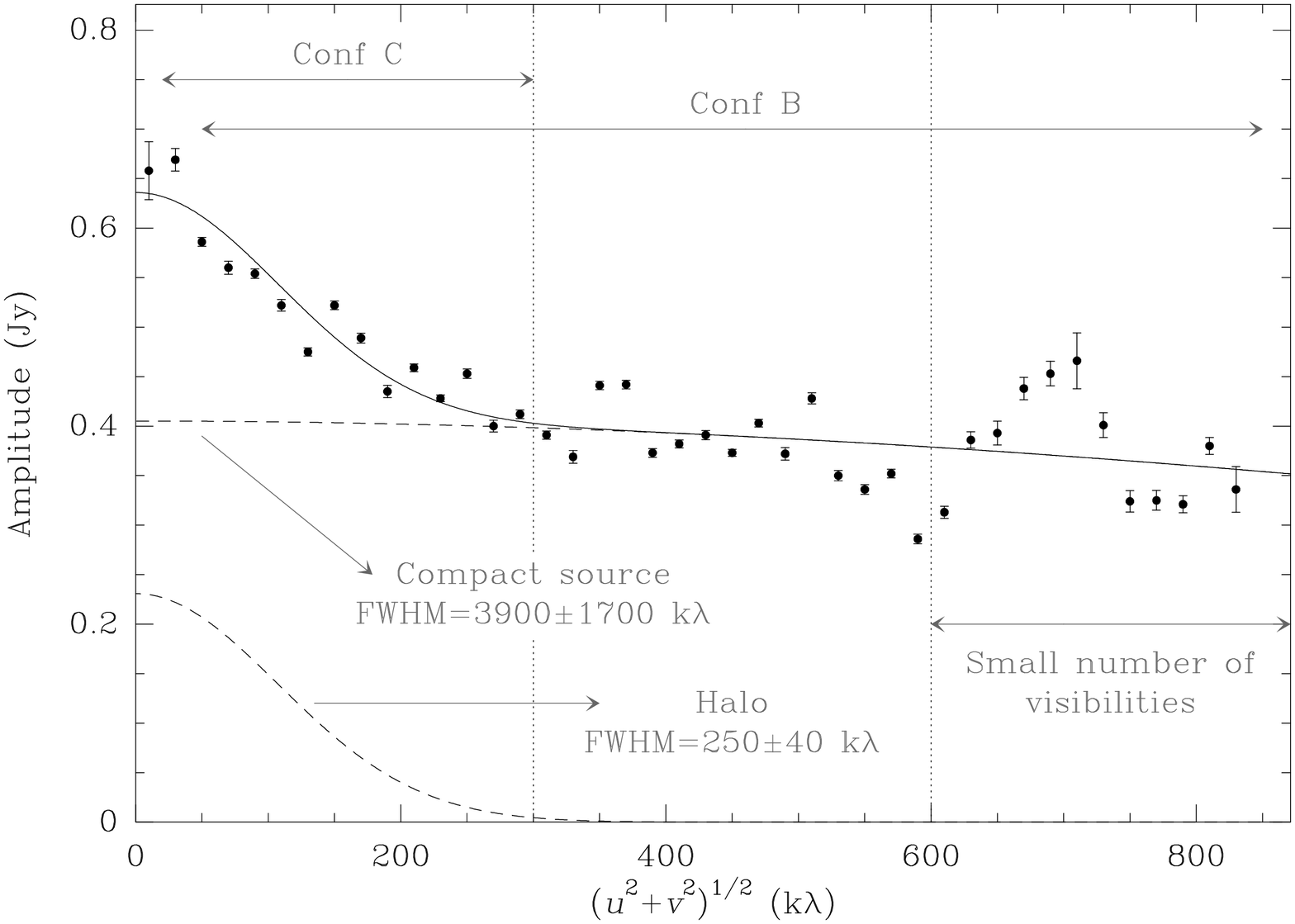}
\caption{Amplitude of the visibility for the continuum emission as a function of
the $uv$-distance.
The amplitude has been averaged out in rings of 20~\klam{} width with the 
\textsc{Miriad} routine \textsc{uvamp}.
Error bars give the typical deviation of the amplitude for each ring.
The data can be fitted with two Gaussians (dashed black curves).
Their sum has been plotted as the solid black curve.}
\label{fig:f2}
\end{figure}

\begin{table*}
\begin{minipage}{\textwidth}
\caption{Characterisation of the moment 0 maps of the most intense lines 
with natural weighting (\texttt{robust=2})}
\label{tab:table3}
\begin{tabular}{c@{}c@{}c@{}c@{}c@{\hspace{2ex}}c@{\hspace{2ex}}c@{\hspace{2ex}}c@{\hspace{2ex}}c@{\hspace{2ex}}c@{\hspace{2ex}}c}
\hline
     &       & \multicolumn{3}{c}{\hrulefill\hspace{1ex}Observations\hspace{1ex}\hrulefill} & \multicolumn{3}{c}{\hrulefill\hspace{1ex}Gaussian fit\hspace{1ex}\hrulefill} & \multicolumn{2}{c}{\hrulefill\hspace{1ex}PSF\hspace{1ex}\hrulefill} & \\
Line & Conf. & $F_\subscript{int}$ & $F_\subscript{max}$    & Peak                              & Centre & Size & P.A.                                                         & Size            & P.A.                                              & BR\\
     &       & (Jy~\kms)         & (Jy~beam$^{-1}$~\kms) & (mas)                             & (mas) & (arcsec$^2$) & ($\degr$)                                          & (arcsec$^2$) & ($\degr$)                                         & \\
\hline
SiS($0;14-13$) & C & $1710\pm 140$ & $1200\pm 25$  & $(90,30)$ & $(78,1)$ & $0.97\times0.79$ & $-85$ & $0.80\times0.65$ & $-78$ & 1.2\\
               & B & $740\pm 70$ & $200\pm 8$ & $(-30,30)$  & $(-18,0)$ & $0.58\times0.48$ & $80$ & $0.28\times0.24$ & $62$ & 1.0\\
SiS($1;14-13$) & B & $54\pm 4$ & $33.6\pm 0.8$ &  $(0,-20)$ & $(-12,-11)$ & $0.35\times0.31$ & $47$ & $0.28\times0.23$ & $50$ & 1.2\\
H$^{13}$CN($0;3-2$) & BC & $600\pm 90$ & $50.1\pm 2.3$ & $(30,30)$ & $(10,10)$ & $0.71\times0.54$ & $70$ & $0.41\times0.36$ & $65$  & 0.8\\
                   &   &  &  &  & $(29,20)$ & $2.37\times1.97$ & $-30$ &  &   & \\
                   & B & $47\pm 9$ & $13.0\pm 1.3$ & $(-15,-20)$ & $(-98,5)$ & $0.78\times0.51$ & $70$ & $0.28\times0.23$ & $63$ & 1.1\\
H$^{13}$CN($\nu_2^{1e};3-2$) & B & $13.1\pm 2.0$ & $9.0\pm 0.7$ & $(-15,-20)$ & $(-22,-11)$ & $0.36\times0.34$ & $49$ & $0.28\times0.23$ & $63$  & 0.8\\
H$^{13}$CN($\nu_2^{1f};3-2$) & B & $13.7\pm 1.7$ & $10.6\pm 0.5$ & $(-45,-20)$ & $(-26,-12)$ & $0.35\times0.29$ & $73$ & $0.28\times0.23$ & $58$  & 0.9\\
SiO($0;6-5$) & BC & $390\pm 60$ & $44.7\pm 1.8$ & $(-60,-50)$ & $(22,-19)$ & $1.02\times0.85$ & $29$ & $0.44\times0.40$ & $75$  & 0.7\\
             &    & & & & $(-78,174)$ & $3.31\times2.28$ & $39$ & \\
             & B & $52\pm 7$ & $11.4\pm 0.9$ & $(-135,-170)$ & $(-76,-64)$ & $0.69\times0.53$ & $10$ & $0.28\times0.25$ & $63$ & 0.6\\
SiO($1;6-5$) & BC & $10.1\pm 2.1$ & $5.2\pm 0.6$ & $(-15,-50)$ & $(137,-4)$ & $0.99\times0.70$ & $52$ & $0.52\times0.39$ & $37$  & 0.5\\
             & B & $4.0\pm 1.0$ & $2.6\pm 0.5$ & $(-75,-10)$ & $(-11,-63)$ & $0.49\times0.36$ & $-13$ & $0.29\times0.22$ & $61$ & ---\\
SiC$_2$($0;11_{2,10}-10_{2,9}$) & BC & $31.3\pm 2.2$ & $6.7\pm 0.5$ & $(0,-50)$ & $(-9,-52)$ & $0.57\times0.49$ & $50$ & $0.53\times0.39$ & $36$  & 1.1\\
                              &    & & & & $(114,102)$ & $2.10\times1.94$ & $-6$ & \\
                              & B & $4.6\pm 1.1$ & $3.1\pm0.5$ & $(-15,-60)$ & $(-7,-83)$ & $0.44\times0.38$ & $70$ & $0.29\times0.23$ & $61$ & 1.2\\
\parbox{25ex}{\centering SiC$_2$($\nu_3;11_{5,7}-10_{5,6}$)\\ 
+\\ 
SiC$_2$($\nu_3;11_{5,6}-10_{5,5}$)} & B & $2.9\pm 0.8$ & $2.9\pm 0.5$ & $(-15,-20)$ & $(1,8)$ & $0.37\times0.35$ & $-57$ & $0.29\times0.23$ & $61$ & 1.2\\
\hline
\end{tabular}
\medskip
\newline
From left to right, the columns contain the label of each molecular line, the 
CARMA configuration (B, BC, or C), the integrated flux density over the 
$3\sigma$ level, the integrated peak emission and its position with respect to 
the star (RA and DEC), the centre of the Gaussian used in the fit 
with respect to the star (RA and DEC), its FWHM and position angle, 
the HPBW and position angle of the PSF, and the blue-red factor defined as the 
ratio of the flux density of the blue-shifted emission to that of the 
red-shifted one with respect to the systemic LSR velocity.
The observed distributions have been fitted adopting one or two Gaussians with
the routine \textsc{imfit}.
We have assumed as reasonable fits those for which the fitting function departs 
from the distribution in less than $3\sigma$.
When the fit to the observed brightness distribution must be performed with two 
Gaussians, the parameters of the second Gaussian are arranged in the following 
line.
All the positions are expressed as offsets with respect to the position of the 
central star (Section~\ref{sec:source.position}).
The uncertainties of most of the axis of the Gaussians are smaller than 10~per~cent 
although they reach 30~per~cent in few cases.
The uncertainty of the position angles is usually smaller than 10\degr.
The blue-red factor of line SiO($v=1,J=6-5$) in B configuration
cannot be estimated because the blue-shifted emission is too weak.
\end{minipage}
\end{table*}

In Figure~\ref{fig:f2} we have plotted the amplitude of the visibility as a 
function of the $uv$-distance.
A simple inspection leads to the identification of two different regimes below 
and over $\simeq 200-250$~\klam, where the gradient of the amplitude changes 
significantly.
The data set can be fitted with two Gaussians centred in the origin of the $uv$ 
plane.
Their FWHMs are $250\pm 40$ and $3900\pm 1700$~\klam, i.e., $0.73\pm 0.12$ and 
$0.05\pm 0.02$~arcsec in the image plane, respectively.
The average amplitudes over 600~\klam{} are noisier than the rest of the data 
set probably as a consequence of the scarcity of long baselines in B 
configuration.
Therefore, to first order, the continuum emission comes from a compact source 
embedded in a halo.

\subsection{Molecular emission}
\label{sec:molecules}

\begin{figure*}
\includegraphics[width=\textwidth]{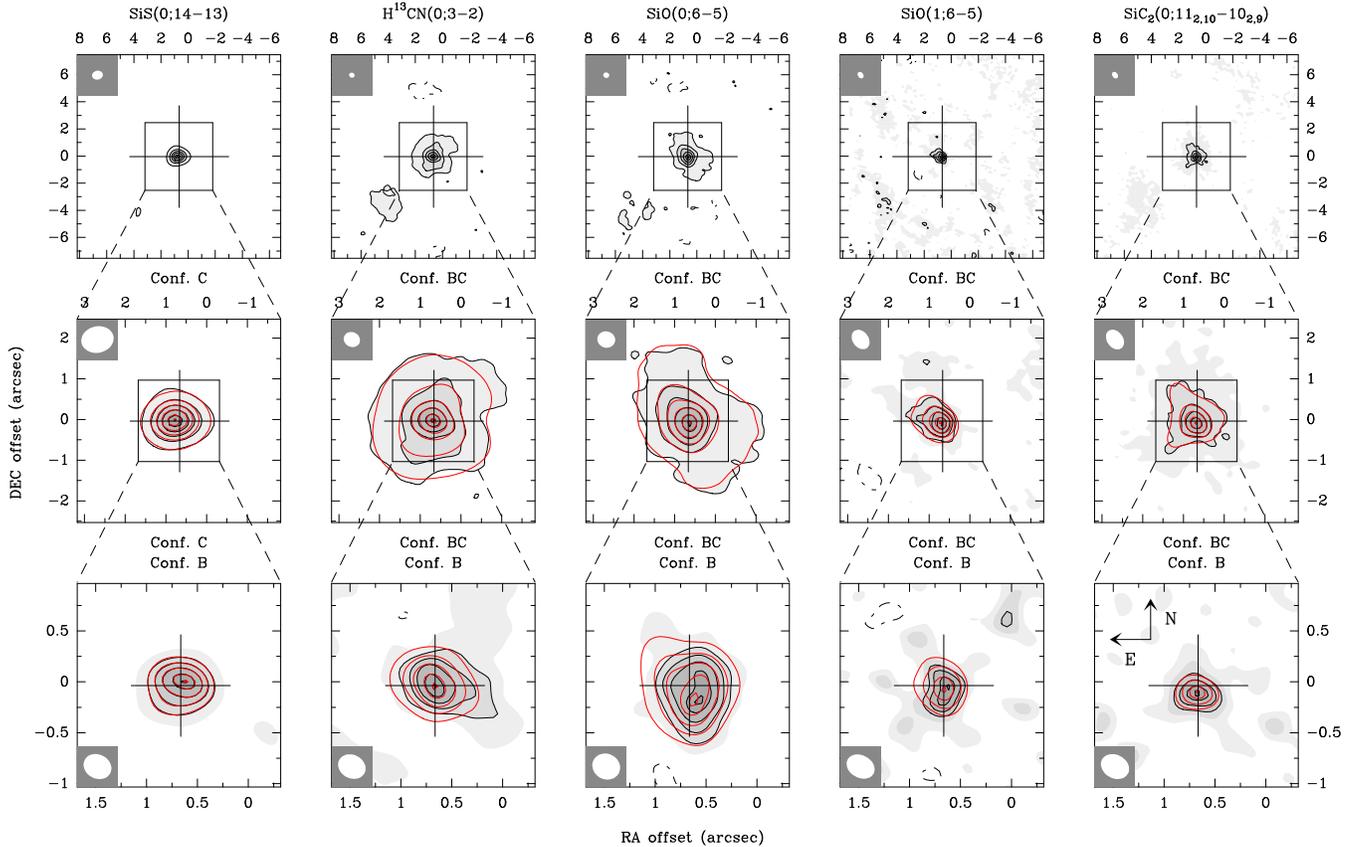}
\caption{Moment 0 maps of the most intense observed lines:
SiS($v=0,J=14-13$), H$^{13}$CN($v=0,J=3-2$), SiO($v=0,J=6-5$), SiO($v=1,J=6-5$), 
and SiC$_2$($v=0,J_{K_a K_b}=11_{2,10}-10_{2,9}$).
The top and middle rows show the maps for
BC configuration (except for the SiS line) while the bottom row 
includes the maps of the observations in B configuration.
In the case of SiS($v=0,J=14-13$), the map in C configuration has been included
instead that in BC configuration 
because of an incompatibility detected
between the observations in B and C configurations
(see Sections~\ref{sec:molecules} and \ref{sec:discussion.sis}, and
Table~\ref{tab:table4}).
Each column contains the brightness distributions of a single line.
All the distributions have been mapped with natural weighting
(\texttt{robust=2}).
The coordinates in the maps are the offsets with respect to the centre of 
phases located at 
$\alpha(\textnormal{J2000})=09\textnormal{h}47\textnormal{m}57.39\textnormal{s}$ and
$\delta(\textnormal{J2000})=+13\degr16\arcmin43\farcs90$.
The black crosses indicate the position of the star 
(Section~\ref{sec:source.position}).
The positive contours for each insert are at levels 10, 30, 50, 70, 90,
and 99~per~cent of the peak emission.
The highest levels are always present.
The minimum level depends on the map.
The only plotted negative contour is at a level equal to the opposite of the 
lowest positive contour.
See Table~\ref{tab:table3} for the peak emissions.
The red contours in the middle and lower rows are the fits to the observed
maps performed with our code (see Section~\ref{sec:modelling}).}
\label{fig:f3}
\end{figure*}

The correlator setup allowed for the simultaneous detection of several molecular
lines through a band width coverage of $\simeq 6.4$~GHz.
The spectrum of both sidebands can be seen in Figure~\ref{fig:f1}.
We have found 25 molecular lines over the $3\sigma$ detection level. 
Twenty of them have been identified.
Four unidentified lines have also been detected in the molecular survey ranging 
from 255.3 to 274.8~GHz carried out with ALMA and partially presented by 
\citet{cernicharo_2013}.
Table~\ref{tab:table2} contains their frequencies and the parameters derived 
from a Gaussian-line fitting procedure.
Those 20 identified lines have been observed in B configuration and 5 of them 
also in C configuration.

Figure~\ref{fig:f3} shows the moment 0 maps of some of the identified lines in 
B and the jointly combined BC configurations calculated with natural weighting
(\texttt{robust=2}).
The observed distributions are roughly fitted with one or two elliptical Gaussians 
using the \textsc{Miriad} routine \textsc{imfit}. 
We have assumed as reasonable fits those with less than $3\sigma$ residuals at 
any position inside the fitting region. 
Table~\ref{tab:table3} contains a summary of these fits. 
In Figure~\ref{fig:f4} we have plotted the red- and blue-shifted emission of 
the H$^{13}$CN and SiO lines in the vibrational ground state.

In order to compare the flux measurements in B and C configurations for the
lines observed in both, we have determined the average flux density for each 
configuration in the region of the $uv$ plane where the set of $uv$ distances 
for both configurations overlap ($200-300$~\klam). 
The C/B ratio between fluxes from C and B configuration ranges from 0.8 to 1.5 
for most of them (see Table~\ref{tab:table4}). 
However, this ratio raises to $\simeq 9$ for line SiS($v=0,J=14-13$). 
This variation is not due to uncertainties in the flux calibration and,
therefore, represents an actual evolution of the line emission (see 
Section~\ref{sec:discussion.sis}).

\subsubsection{SiS}
\label{sec:molecules.sis}

The brightness distributions of line SiS$(v=0,J=14-13)$ in C and B 
configurations show circular symmetry (Figure~\ref{fig:f3}). 
Their sizes are smaller than expected for a line supposed to display extended 
emission.
The peak emission is located roughly on the star in the naturally weighted
maps although an offset due NW might exist
(Table~\ref{tab:table3}).
However, it is offset with respect to the star by 
$\simeq 70\pm 40$~mas due NW in the uniformly weighted (\texttt{robust=-1}) 
B configuration map (not shown;
HPBW~$=0.27\times0.20$~arcsec$^2$, P.A.~$=87\degr$).
The emission in this map is barely resolved and required a two-Gaussian fit. 
These two Gaussians are centred at the peak emission and at the SE of the star 
position. 
The blue-red factor, defined as the ratio of the flux densities from the blue- 
and red-shifted emission, is 1.2 and 1.0 in the naturally weighted maps
in the C and B configurations, respectively (see Table~\ref{tab:table3}). 

The emission of the vibrationally excited line SiS($v=1,J=14-13$) 
(not shown)
is characteristic of a compact source centred at the star position. 
Its blue-red factor is 1.2.

\subsubsection{H$^{13}$CN}
\label{sec:molecules.h13cn}

The brightness distribution of line H$^{13}$CN$(v=0,J=3-2)$ in BC configuration
shows an overall extended emission more prominent along the N-S direction 
at 2~arcsec scale (Figure~\ref{fig:f3}). 
At smaller scales ($\simeq 1$~arcsec) there is a possible E-W elongation in 
the surroundings of the star. 
This is reflected in the two-Gaussian fit we carried out to the emission. 
Figure~\ref{fig:f4} shows that the red-shifted emission in B configuration
mapped with natural weight peaks at the star position while the 
blue-shifted emission displays an elongation roughly along the NE-SW direction.
The uniformly weighted map in B configuration suggests that this elongation
is composed of two clumps, one to the N-NE of the star and the other to the 
S-SW.

The vibrationally excited lines H$^{13}$CN($\nu_2^{1e},J=3-2$) and 
H$^{13}$CN($\nu_2^{1f},J=3-2$) (not shown)
display a compact point-like brightness distribution centred at the star.

\subsubsection{SiO}
\label{sec:molecules.sio}

\begin{figure}
\centering
\includegraphics[width=0.475\textwidth]{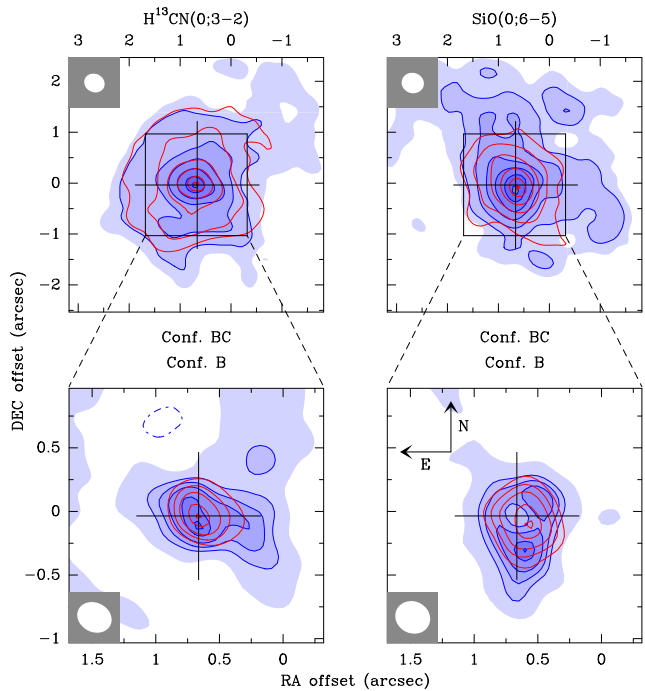}
\caption{Red- and blue-shifted emission (in red contours
and blue contours and colour scale, respectively) of 
lines H$^{13}$CN($v=0,J=3-2$) and SiO($v=0,J=6-5$) in BC (upper inserts) and B 
(lower inserts) configurations.
All the emissions have been mapped assuming natural weighting
(\texttt{robust=2}).
The contours in the maps in BC configuration are at levels $-15$, 15, 30, 50, 
70, 90, and 99~per~cent of the peak emission.
They are at $-40$, 40, 55, 70, 90, and 99~per~cent of the peak emission for maps in B 
configuration.
The colour scale in the lower right insert reveals an emission deficit around 
the star in the blue-shifted emission map.}
\label{fig:f4}
\end{figure}

Figure~\ref{fig:f3} shows the SiO($v=0,J=6-5$) emission mainly extended and with
an overall preferred NE-SW direction at 3~arcsec scales. 
At these scales, the blue-shifted emission is preferentially seen towards the 
NE, while the red-shifted emission is elongated towards the SW, evidencing a 
possible bipolar outflow along this direction, although with a severe overlap 
between the red- and blue-shifted emission towards the central region 
(Figure~\ref{fig:f4}).
 
The peak emission of the distribution of line SiO$(v=0,J=6-5)$ in the high 
angular resolution maps (with a $\simeq 0.25$~arcsec beam) is shifted to the SW. 
The red-shifted emission is single peaked 
while the blue-shifted emission shows multiple peaks forming an arc-like 
feature surrounding the position of the red-shifted peak. 

In good agreement with the other vibrationally excited molecular transitions, 
SiO$(v=1,J=6-5)$ is a compact source peaking at the $5\sigma$ level.
A Gaussian fit to its BC configuration emission gives a somewhat elongated 
source with a P.A.~$\simeq 50\degr$. 

\subsubsection{SiC$_2$}
\label{sec:molecules.sic2}

The SiC$_2(v=0,J_{K_a K_b}=11_{2,10}-10_{2,9})$ emission is mostly circular at 
scales of 1~arcsec but it seems to be elongated along the NE and SE 
directions at scales of 2~arcsec.

The emission of the vibrationally excited lines in B configuration 
(not shown) is compact and centred on the star position.

\begin{table}
\caption{Flux density in the overlap region of the $uv$ plane}
\label{tab:table4}
\begin{tabular}{cccc}
\hline
Source & Conf. B & Conf. C & Factor C/B\\
\hline
Continuum                       & $405\pm 5$ &  $389\pm 3$  & $1.0$\\
H$^{13}$CN($0;3-2$)              & $470\pm 4$ &  $724\pm 20$ & $1.5$\\
SiO($0;6-5$)                    & $466\pm 5$ &  $565\pm 20$ & $1.2$\\
SiO($1;6-5$)                    & $421\pm 5$ &  $383\pm 20$ & $0.9$\\
SiC$_2$($0;11_{2,10}-10_{2,9}$)    & $401\pm 5$ &  $309\pm 20$ & $0.8$\\
SiS($0;14-13$)                  & $925\pm 4$ & $8100\pm 40$ & $8.8$\\
\hline
\end{tabular}
\medskip
\newline
The first column contains the source (molecular line or continuum).
The second and third columns account for the average amplitudes expressed in 
mJy in the overlap region of the $uv$ plane derived from the observations in B 
and C configurations.
The fourth column is devoted to the ratio of the amplitudes in C configuration 
to B in the overlap range.
Most of the ratios depart from 1 in no more than $\simeq 50$~per~cent, typical 
for compatible observations of different configurations.
The ratio for line SiS($v=0,J=14-13$) is much larger suggesting an unavoidable
incompatibility between the observations in B and C configurations.
\end{table}

\section{Fitting the molecular emission}
\label{sec:modelling}

The observations presented in Section~\ref{sec:molecules} evidence the 
existence of structure in the emission of each molecular line not 
mapped to date.
Deviations from spherical symmetry could be due to particular features 
or global asymmetries either 
in the abundance with respect to H$_2$
and/or excitation temperature distributions of each molecule,
or in magnitudes describing the gas component of the envelope
as the H$_2$ density and/or 
the expansion velocity field. 
Other structures such as sets of clumps have not been considered
since these clumps would be substantially smaller than the PSF of our 
observations and their effect could not be distinguished from that of an
equivalent continuous structure.
In order to describe the molecular content of the dust formation zone in 
\irc, we have explored the spatial dependence of the abundance with respect 
to H$_2$ and the vibrational temperature for each molecule. 
The analysis of the magnitudes that describe
the gas (H$_2$ density, expansion velocity)
has been done after fitting
the molecular observations by looking for shared features 
or asymmetries in all the abundance distributions.

The fits to the emission of the molecular lines have been performed with an 
improved 3D version of the 1D radiation transfer
code developed by \citet{fonfria_2008}, which 
numerically solves the radiation transfer equation in
a spherically symmetric
circumstellar envelope composed of molecular gas and dust.
The new version of the 
code is able to deal with diatomic, linear, and symmetric top
molecules in addition to asymmetric top molecules after 
diagonalising the Watson's Hamiltonian in its A-reduction I$^r$ representation,
suitable for the case of SiC$_2$ \citep{fonfria_2008,fonfria_2011,muller_2012}.
The statistical equilibrium equations \textit{are not solved}.
The populations of the molecular levels are computed from the
rotational and vibrational temperature distributions that are input data
as well as the molecular abundance with respect to H$_2$, 
the H$_2$ density profile, and the
expansion and turbulence velocity fields
(see Section~\ref{sec:appendix.program} of Appendix~\ref{sec:appendix}
for a deeper description of the program).
Our code can also reproduce maser emission assuming a 
negative excitation temperature in the maser emitting region to get the 
required population inversion.
The contribution of the maser emitting volumes to the total synthetic
emission is calculated in the same way and with the same mathematical 
expressions than for the thermal emitting volumes.
Thus, the velocity coherence has been taken into account and the code 
derives the exact solutions for the maser emission
within the typical uncertainty $\lesssim 1$~per~cent.
This code was developed to solve
the radiation transfer equation
for any gas expansion velocity field adopted,
allowing us to derive the emission of any region
of an asymmetric CSE. 
Thus, it is particularly useful to analyse the innermost envelope of
\irc, where the velocity gradient is similar to the line widths.

The code allows any molecular abundance with respect to H$_2$ and 
rotational and vibrational temperature distributions expressed in spherical 
coordinates.
We identify the $z$ axis of the reference system with the line-of-sight (LoS).
The polar angle, $\theta$, is measured from behind the star towards the Earth 
and the axial angle (or position angle), $\varphi$, from N to E.
The code calculates the emission for several axial angles and performs a linear
interpolation on $\varphi$ for any other axial angle
(see Section~\ref{sec:appendix.gridding} of Appendix~\ref{sec:appendix}
for a description of the
gridding and sampling of the physical and chemical magnitudes).
A realistic coverage of the $uv$ plane has been achieved by replacing the 
calibrated visibilities in the \textsc{Miriad} data files with those of the 
synthetic emission using the \textsc{Miriad}'s routine \textsc{uvmodel}.
This method applies on the synthetic emission the same constraints 
imposed by the array configuration on
the actual brightness distribution.

The performance of our code was tested in 1D against
($i$) simple scenarios, whose solutions have been calculated analytically or
numerically through independent ad-hoc codes, and
($ii$) the non-local, non-LTE radiation transfer code 
developed by \citet{daniel_2008} for
static and expanding envelopes which emits optically thin and thick lines
produced under LTE, with positive
excitation temperatures depending on the distance
to the star, or with negative excitation temperatures
to model maser emission (see Appendix~\ref{sec:appendix.benchmark}).
The agreement between the results was always better than 1~per~cent
of the maximum emission and usually significantly better than 
0.1~per~cent.
Further comparison 
with analytical and numerical independent ad-hoc codes to simple
2D and 3D scenarios gave emission differences always
smaller than 0.5~per~cent.

\begin{table}
\caption{Fixed parameters used in the molecular fits}
\label{tab:table5}
\begin{tabular}{cccc}
\hline
Parameter & Value & Units & Ref. \\
\hline
$D$ & 123 & pc & 2\\
$\dot M$ & $2.0\times 10^{-5}$ & M$_\odot$~yr$^{-1}$ & 3\\
$T_\star$ & 2330 & K & 4\\
$\alpha_\star$ & 0.02 & arcsec & 3\\
$\rstar$ & $3.7\times 10^{13}$ & cm & \\
$\rin$ & 5 & $\rstar$ & 5\\
$\rout$ & 20 & $\rstar$ & 5\\
$v_\textnormal{\tiny exp}(\textnormal{Region I})$ & 5 & \kms & 5\\
$v_\textnormal{\tiny exp}(\textnormal{Region II})$ & 11 & \kms & 5\\
$v_\textnormal{\tiny exp}(\textnormal{Region III})$ & 14.5 & \kms & 5\\
$\Delta v(r=\rstar)$ & 5 & \kms & 3\\
$\Delta v(r>\rin)$ & 1 & \kms & 3\\
$T_\subscript{rot}(\textnormal{H}^{13}\textnormal{CN};\textnormal{Regions I-II})$ & $T_\star(\rstar/r)^{0.58}$ &  & 5\\
$T_\subscript{rot}(\textnormal{H}^{13}\textnormal{CN};\textnormal{Region III})$ & $T_\subscript{out}(\rout/r)$ &  & 5\\
$T_\subscript{rot}(\textnormal{SiO};\textnormal{Regions I-III})$ & $T_\star(\rstar/r)^{0.55}$ &  & 3\\
$T_\subscript{rot}(\textnormal{SiC}_2;\textnormal{Regions I-II})$ & $T_\star(\rstar/r)^{0.81}$ &  & 6\\
$T_\subscript{rot}(\textnormal{SiC}_2;\textnormal{Region III})$ & $T_\subscript{out}$ &  & 6\\
$\tau_\textnormal{\tiny dust}(257~\textnormal{GHz})$ & $1.5\times 10^{-3}$ &  & 1\\
$T_\textnormal{\tiny dust}(r=\rin)$ & 830 & K & 1 \\
$\gamma_\textnormal{\tiny dust}$ & 0.375 &  & 3\\
\hline
\end{tabular}
\medskip
\newline
$D$: distance to the star; 
$\dot M$: mass-loss rate; 
$T_\star$: effective stellar temperature; 
$\alpha_\star$: angular stellar radius; 
$\rstar$: linear stellar radius;
$\rin$ and $\rout$: positions of the inner and outer acceleration zones;
$v_\textnormal{\tiny exp}$: expansion velocity field;
$\Delta v$: line width ($\propto\rstar/r$);
$T_\subscript{rot}$: rotational temperature;
$T_\subscript{out}=T_\subscript{rot}(\rout)$ for each molecule;
$\tau_\textnormal{\tiny dust}$: optical depth of dust;
$T_\textnormal{\tiny dust}$ and $\gamma_\textnormal{\tiny dust}$: temperature of 
the dust grains at the inner acceleration shell and exponent of the power-law
$T_\textnormal{\tiny dust}(\rin/r)^{\gamma_\textnormal{\tiny dust}}$.
The temperature of the dust grains at $\rin$ and the 
optical depth at 257~GHz have been estimated by fitting the observations 
carried out by \textit{ISO}/SWS in the mid-infrared range \citep{cernicharo_1999} 
keeping fixed $\gamma_\textnormal{\tiny dust}$.
It is assumed that there is no dust in Region I.
(1) This work
(2) \citet{groenewegen_2012}
(3) \citet{agundez_2012}
(4) \citet{ridgway_1988}
(5) \citet{fonfria_2008}
(6) \citet{cernicharo_2010}
\end{table}

We fitted the observed red-shifted, blue-shifted, and moment 0 maps.
Several parameters were kept fixed in our calculations and their
values were taken from the literature (Table~\ref{tab:table5}).
The envelope is divided into three Regions (I, II, and III, outwards from the 
star) separated by two acceleration zones at 5 and $20~\rstar$.
The adopted expansion velocity field, $v_\subscript{exp}$,
is a spherically symmetric step function of the radial
distance to the star, taking a constant value of
5, 11, and 14.5~\kms{} in Regions I, II, and III, respectively.
The H$_2$ density profile was also spherically symmetric and
$\propto r^{-2}v_\subscript{exp}^{-1}$.
We assumed a line width due to turbulence and the gas kinetic temperature
at the stellar surface of 5~\kms, decreasing down to 1~\kms{} 
at the inner acceleration zone following a power-law, 
and remains constant outwards.
The rotational temperatures of all the observed molecules 
were assumed to follow a power-law depending on the distance to the 
star.
They were taken from previous work and kept fixed during the fitting procedure
since in most cases
only one line per vibrational state are available
in the observed data set.
Regarding the spectroscopic data,
the rest frequencies of the observed lines were taken from The CDMS 
Catalogue.
We used the spectroscopic constants by \citet{muller_2007}, 
\citet{maki_2000}, \citet{sanz_2003}, and \citet{muller_2012} and 
\citet{izuha_1994} for SiS, H$^{13}$CN, SiO, and SiC$_2$, respectively.
The dust was assumed to be composed of amorphous carbon.
The dust optical refractive indexes needed to calculate the dust emission at 
the observed frequencies ($n\simeq 2.6$ and $k\simeq 0.071$ at 257~GHz) were
estimated by linear extrapolation from the results by \citet{suh_2000} at 
wavelengths up to 1~mm.

The parameters for each molecular line varied during the fitting procedure
were the abundance with respect to
H$_2$ and the vibrational temperature at 5, 20, and 50~\rstar{} for several 
axial angles.
Those magnitudes in Region I were allowed to depend on the axial angles but
they did not depend on the distance to
the star, since the angular resolution of our observations is not high enough
to resolve this Region.
All the parameters were adopted to be independent of the polar angle 
in front of and behind the star reflecting the low spectral resolution
of the observations but they can be different in both hemispheres, i.e., 
the derived 3D abundance and vibrational temperature distributions are 
composed of two 2D distributions depending on the distance to the star and
the axial angle.
The number of axial angles used ranged from 9 to 12.
The parameters were found to be statistically
significant within a confidence interval of 95~per~cent or larger 
after applying a multivariate statistical significance test,
provided the axial width of a structure is larger than 35\degr{} in Region I
and 15\degr{} in Region III.

The procedure followed to fit the observed maps  
starts with spherically symmetric abundance and temperature 
distributions adopted from the literature.
The red-shifted, blue-shifted, and moment 0
emissions of all the lines of a given molecule are fitted at the same time.
For each line, maps in B configuration are fitted first.
The derived abundance and temperature distributions are used as initial 
conditions to fit the maps in BC configuration.

The fits have been performed several times adopting different starting points
in the space of parameters in order to look for substantially different minima
of the $\chi^2$ function.
The results of this search suggest that the solutions to the fits
are not degenerated.
However, the space of parameters is huge and a complete exploration would
required a very large computational effort.
Thus, we might have overlooked other different solutions
that also describe the observations.

The synthetic and observed brightness distributions
usually differ in less than $1\sigma$ and 
always in less than $2\sigma$, 
where $\sigma$ is the rms noise level of the observed image.
The uncertainties of the abundances and the vibrational temperatures
have been estimated by varying these magnitudes until the differences
between the synthetic and the observed brightness
distributions change in $1\sigma$ at any position.
The error of the abundance and the vibrational temperature
is between 25~per~cent and a factor of 2.
The parameters in Region I show the largest relative errors in both
magnitudes.

\subsection{SiS}
\label{sec:modelling.sis}

\begin{figure}
\includegraphics[width=0.475\textwidth]{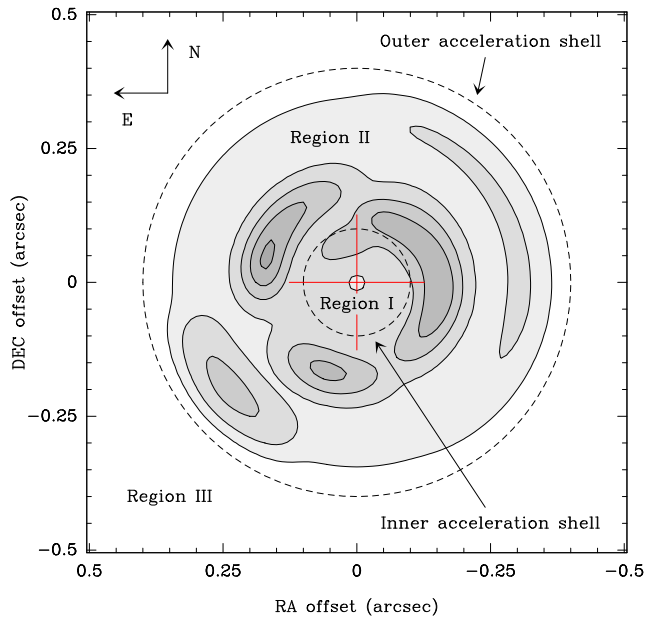}
\caption{Moment 0 emission fitting the observed brightness distribution of line 
SiS($v=0,J=14-13$).
The brightness distribution has been computed convolving the synthetic emission 
with a Gaussian beam of HPBW~$=50$~mas.
The red cross and the dashed circles mark the positions of the central star and 
the acceleration zones at 5 and $20~\rstar$, respectively.
The plotted contours are at 3.0, 6.5, 20, 50, and 90~per~cent of the peak 
emission ($\simeq 29.8$~Jy~beam$^{-1}$).}
\label{fig:f5}
\end{figure}

The fits to the observed emission of the SiS lines have been performed assuming
a circularly symmetric abundance distribution equal to $5.0\times 10^{-6}$ in 
Region I, which decreases linearly down to $1.5\times 10^{-6}$ and 
$1.3\times 10^{-6}$ at the outer acceleration shell and 50~\rstar,
respectively, remaining constant outwards 
(Fonfr\'ia et al., in preparation).
The adopted rotational temperature for the vibrational states $v=0$ and 1 has 
been $2330(\rstar/r)^{0.6}$~K and the vibrational temperature,
$1550(\rstar/r)^{0.6}$~K for band $v=1-0$ and 
$1300(\rstar/r)^{0.6}$~K for band $v=2-1$ and those involving higher
vibrational states. 
Under these conditions, the thermal emission accounts only for a maximum of 
30~per~cent of the observed flux density of line SiS($v=0,J=14-13$).
More complex models such as a point-like or an arc-like maser emitting regions 
and a thermal emitting halo can barely reproduce the region around the peak 
emission of this line but fail to fit the rest of the distribution.
The synthetic emission deviates from the observed in more than $3\sigma$ in a 
significant fraction of the mapped area.
The observed emission can be properly fitted with
a set of arc-like maser emitting regions 
composed of gas with an expansion velocity projected upon the LoS between 
$-12$ and 12~\kms{} and a weak circularly symmetric thermal contribution, mostly
coming from Regions I and II (Figure~\ref{fig:f5}).
The radii of the arcs in the plane of the sky
range from 4 to $15~\rstar$ (i.e., an angular distance of
80 to 300~mas) and their fluxes decrease with the distance to the star.
Most of the emission comes from the northern hemisphere and is produced close 
to the inner acceleration shell.
Maser emission depends largely on the size and depth of the emitting region.
None the less, the SiS maser emitting structure is unresolved
with the present observations and the depth of the arcs cannot be reliably
estimated.
This prevents us from deriving the SiS density and excitation temperature in 
the proposed structure.

Line SiS($v=1,J=14-13$) cannot be reproduced either assuming thermal emission
with the adopted abundance and temperature distributions.
The synthetic emission is about 60~per~cent smaller than the observed brightness
distribution.
No variation in the vibrational temperature significantly improves the 
quality of the fit.
We address this issue in Section~\ref{sec:discussion.sis}.

\subsection{H$^{13}$CN}
\label{sec:modelling.h13cn}

\begin{table}
\caption{Average abundance profiles, $\langle x\rangle_{\varphi\theta}$}
\label{tab:table6}
\begin{tabular}{c@{\hspace{9ex}}c@{\hspace{9ex}}c@{\hspace{9ex}}c}
\hline
Radius & H$^{13}$CN & SiO & SiC$_2$\\
($\rstar$) & $(\times 10^{-7})$ & $(\times 10^{-7})$ & $(\times 10^{-7})$\\
\hline
\multicolumn{4}{c}{\textit{Moment 0 emission}}\\
\hline
$1-5^-$& $8 \pm 3$      & $\lesssim 0.20$ &  $8\pm 5$\\
$5^+$  & $8 \pm 3$      & $2.4 \pm 0.8$ &  $8\pm 5$\\
20     & $3.3 \pm 1.0$  & $3.3\pm 1.0$  &  $0.8 \pm 0.7$\\
50     & $2.5 \pm 0.8$  & $1.8\pm 0.9$  &  $3.4\pm 1.7$\\
\hline
\multicolumn{4}{c}{\textit{Red-shifted emission}}\\
\hline
$1-5^-$& $9 \pm 4$      & $\lesssim 0.25$  &  $6\pm 4$\\
$5^+$  & $9 \pm 4$      & $3.9\pm 1.7$     &  $6\pm 4$\\
20     & $3.6 \pm 1.2$  & $4.9\pm 1.3$     &  $0.9\pm 1.1$\\
50     & $3.5 \pm 1.3$  & $2.4\pm 1.7$     &  $3.8\pm 2.3$\\
\hline
\multicolumn{4}{c}{\textit{Blue-shifted emission}}\\
\hline
$1-5^-$& $5.8 \pm 2.2$  & $\lesssim 0.30$  &  $9\pm 6$\\
$5^+$  & $5.8 \pm 2.2$  & $0.9\pm 0.1$  &  $9\pm 6$\\
20     & $2.9 \pm 0.9$  & $1.8\pm 1.0$  &  $0.7\pm 0.5$\\
50     & $1.5 \pm 0.6$  & $1.1\pm 0.6$  &  $3.0\pm 1.8$\\
\hline
\end{tabular}
\medskip
\newline
The abundances have been calculated by averaging the results of the fits on the 
axial and polar angles.
The dispersions are the standard deviation of the data sets.
The superscripts $^+$ and $^-$ indicate the approach to the inner
acceleration shell from outside of Region I and from inside, respectively.
\end{table}

\begin{figure}
\includegraphics[width=0.475\textwidth]{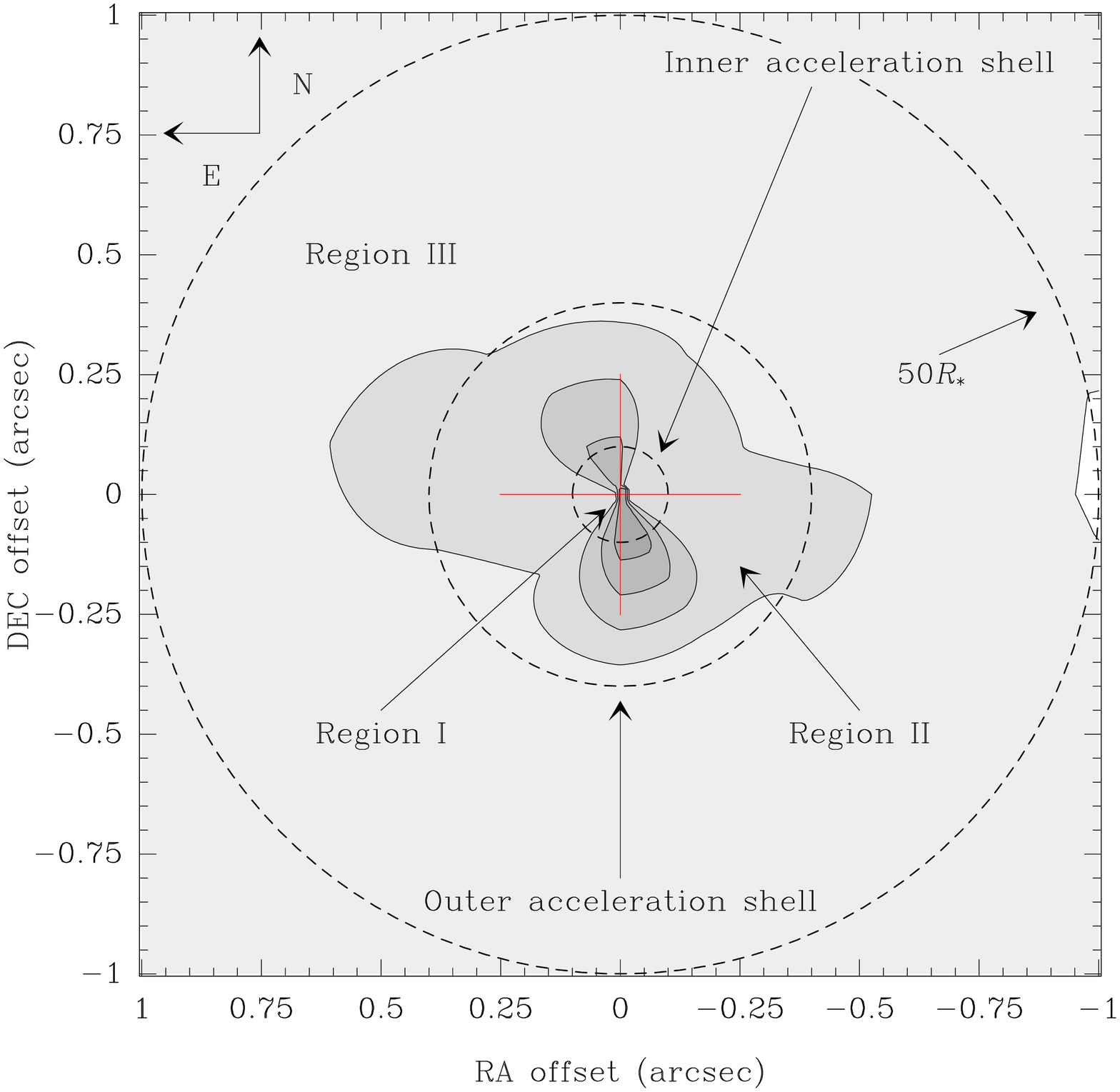}
\caption{H$^{13}$CN abundance with respect to H$_2$ averaged on the polar angle, 
$\langle x\rangle_\theta$.
The contours are at 10, 30, 50, 70, and 90~per~cent of the maximum abundance
($\simeq 1.3\times 10^{-6}$).
The red cross indicates the position of the central star.
The dashed circles are at 5, 20, and $50~\rstar$ from the star.}
\label{fig:f6}
\end{figure}

\begin{figure}
\includegraphics[width=0.475\textwidth]{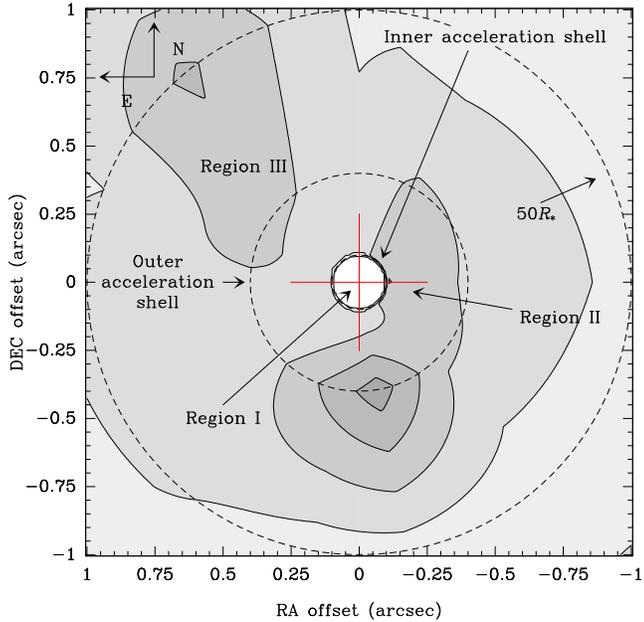}
\caption{SiO abundance with respect to H$_2$ averaged on the polar angle, 
$\langle x\rangle_\theta$.
The contours are at 10, 30, 50, 70, and 90~per~cent of the maximum abundance
($\simeq 6.0\times 10^{-7}$).
The red cross indicates the position of the central star.
The dashed circles are at 5, 20, and $50~\rstar$ from the star.}
\label{fig:f7}
\end{figure}

The H$^{13}$CN abundance averaged on the axial and polar angles, 
$\langle x\rangle_{\varphi\theta}$, decreases by a factor of 
$2-3$ from the stellar 
surface to the outer acceleration zone, remaining nearly constant outwards
(Table~\ref{tab:table6}).
It is larger behind the star in Regions I and II.

The spatial distribution of the H$^{13}$CN abundance (Figure~\ref{fig:f6}) 
reveals the presence of a bipolar structure lying along the NNE-SSW 
direction (P.A.~$\simeq 15\degr$) that extends throughout Region I and
part of Region II.
The abundance in this structure due SSW is about 40~per~cent larger than
due NNE.
In Region II, the H$^{13}$CN abundance is strongly reduced 
by a factor of up to 6.

The H$^{13}$CN lines in the vibrational states $v=1$ can be fitted with a
circularly symmetric vibrational temperature distribution 
of 2300~K in average in Region I.

\subsection{SiO}
\label{sec:modelling.sio}

\begin{figure}
\includegraphics[width=0.475\textwidth]{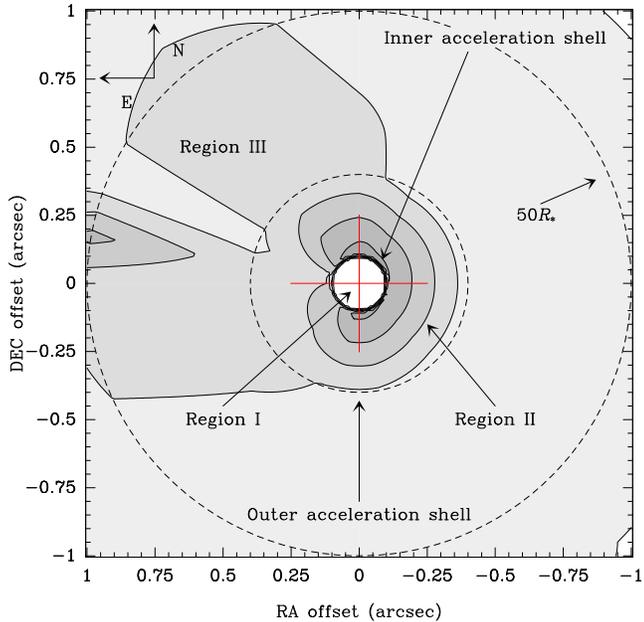}
\caption{Vibrational temperature of SiO averaged on the polar angle, 
$\langle T_\subscript{vib}\rangle_\theta$.
The contours are at 10, 30, 50, 70, and 90~per~cent of the maximum temperature 
($\simeq 2000$~K).
The red cross indicates the position of the central star.
The dashed circles are at 5, 20, and $50~\rstar$ from the star.
The temperature in Region I cannot be reliably determined from our 
observations.}
\label{fig:f8}
\end{figure}

\begin{figure}
\includegraphics[width=0.475\textwidth]{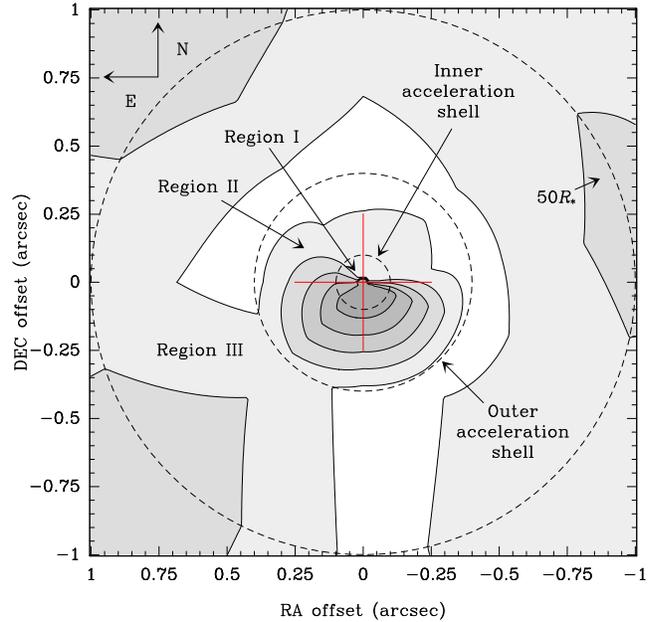}
\caption{SiC$_2$ abundance with respect to H$_2$ averaged on the polar angle, 
$\langle x\rangle_\theta$.
The contours are at 10, 30, 50, 70, and 90~per~cent of the maximum abundance
($\simeq 1.4\times 10^{-6}$).
The red cross indicates the position of the central star.
The dashed circles are at 5, 20, and $50~\rstar$ from the star.}
\label{fig:f9}
\end{figure}

The SiO average abundance $\langle x\rangle_{\varphi\theta}$ in Region I required
to reproduce the emission near the star is $\lesssim 2\times 10^{-8}$, at least 
one order of magnitude smaller than the abundance in Regions II and III
(see Section~\ref{sec:discussion.sio}).
The maximum of the abundance profile is located at the outer acceleration zone.
The fit shows that most of the SiO is located behind the star, specially in 
Region II (Table~\ref{tab:table6}).
The abundance distribution $\langle x\rangle_\theta$ shows two main SiO
abundance peaks (Figure~\ref{fig:f7}).
The strongest is located at the outer acceleration shell due S-SW.
The other peak is found to the NE of the star at $\simeq 50~\rstar$,
apparently behind it.

The average vibrational temperature at the inner and outer acceleration zone
is $\simeq 1750$ and 580~K, respectively.
The spatial distribution of the vibrational temperature displays an elongation 
throughout Region III roughly due NE where it is at least a factor of 2 larger 
than along other directions (Figure~\ref{fig:f8}).

\subsection{SiC$_2$}
\label{sec:modelling.sic2}

The SiC$_2$ average abundance $\langle x\rangle_{\varphi\theta}$ derived from our 
fits decreases throughout Regions I and II and increases outwards 
(Table~\ref{tab:table6}).
The abundance in front of the star is about 2 times larger in Region I than
behind it.

The abundance distribution $\langle x\rangle_\theta$ shows 
a roughly semicircular structure distributed to the S of the star
with a peak emission $5-6$ times larger than due N (Figure~\ref{fig:f9}).
Three increments can be noticed beyond the outer acceleration
shell towards NE, SE, and W 
(P.A.~$\simeq 45\degr$, $135\degr$, and $280\degr$, respectively).
A minimum with an abundance at least a factor of 3 smaller than for other
orientations extends from the outer acceleration shell outwards due S
(P.A.~$\simeq 180\degr$).

The vibrational temperature can be assumed to be circularly symmetric.
It is about 1700~K in Region I and is compatible with a decrease down to about 
200~K at the end of Region II.

\section{Analysis and discussion}
\label{sec:discussion}

This work is based on interferometric observations carried out in two different 
CARMA configurations.
Between both observations there was a period of 14 months.
Following \citet{jenness_2002}, the data in B and C configurations were 
acquired when the phase of \irc{} was 0.8 and 0.5, respectively.
The changes in the radiation field existing in the dust formation zone during a 
pulsation period could largely influence the molecular excitation 
\citep*[e.g.,][]{carlstrom_1990,monnier_1998}.
This probably explains the differences in the continuum and most of the 
molecular line emission between B and C configuration observations 
(Table~\ref{tab:table4}).
However, 
the strong variation of the SiS emission between the two epochs seems to be
related to its maser nature (see Section~\ref{sec:discussion.sis}).

\subsection{Continuum emission}
\label{sec:discussion.continuum}

In Section~\ref{sec:continuum.structure} we proposed that the continuum
emission is composed of a compact source surrounded by a halo.
This structure is compatible with that inferred from previous observations
\citep{lucas_1997,lucas_1999,patel_2009,shinnaga_2009}.

The halo is very probably produced by the dust emission coming from Region II, 
where most of dust is formed and its temperature is still high 
\citep*[e.g.,][]{menshchikov_2001,fonfria_2008}.

The compact source has a size of $50\pm 20$~mas, very similar to the diameter 
of the star 
\citep*[$\simeq 30-50$~mas;][]{ridgway_1988,monnier_2000a,menshchikov_2001}.
The flux density of the star, assumed as a black-body with an effective 
temperature of 2500~K 
\citep{ridgway_1988,ivezic_1996,bergeat_2001,menshchikov_2001},
is $\simeq 150$~mJy at 257~GHz, i.e., a factor of $\simeq 3$ smaller than the 
flux density observed in B and C configurations ($380\pm 60$ and 
$530\pm 80$~mJy).
On the other hand, our code indicates that the dust contribution compared to the
flux density of the star is negligible ($\lesssim 20$~mJy).
Hence, the flux density excess may come from 
($i$) a shell surrounding the star that mostly emits in the mm and probably cm 
wavelength ranges or 
($ii$) from the star itself, if we assume that it does not emit as a black-body.
Both scenarios have already been invoked previously 
\citep{sahai_1989,menten_2006,menten_2012}.

\subsection{SiS}
\label{sec:discussion.sis}

In Section~\ref{sec:modelling.sis} we showed that the emission of line 
SiS($v=0,J=14-13$) can only be explained considering the presence of several 
maser emitting arcs (Figure~\ref{fig:f5}).
The necessity to include maser emission in the model is directly related to
the large ratio of the amplitudes in C to B configurations in the overlap 
range of the $uv$-plane of line SiS($v=0,J=14-13$) (Table~\ref{tab:table4}).
It can be argued that this variation could be produced by errors in the flux 
calibration but, in this case, these errors would be similarly reflected in 
the rest of the observed lines.
However, the amplitude measured for the rest of the lines in both epochs show 
good agreement, indicating that the SiS emission has suffered an intrinsic 
variation directly related to the evolution of the pumping mechanism in that 
period of time.

According to \citet{fonfria_2006}, the most probable pumping mechanism of line 
SiS($v=0,J=14-13$) is produced by the overlap of certain ro-vibrational 
lines of SiS (that we will name excitable lines) with other of C$_2$H$_2$, HCN, 
and their isotopologues with similar frequencies (exciting lines).
The Doppler shift due to the relative motion of separate gas volumes produce a 
frequency shift that favours the overlap of the exciting and the excitable 
lines.
This selective excitation mechanism could invert the population of the levels 
involved in a rotational transition if the intensity of the exciting radiation
and the density of SiS are large enough.
Thus, the maser emitting arcs derived from our fits would match up with regions 
where any of the following magnitudes is larger compared with their 
surroundings:
($i$) the density of SiS,
($ii$) the intensity of the available exciting radiation, or
($iii$) the overlap fraction between the exciting and the excitable lines.
The last case would imply that the expansion velocity field depends on the 
axial angle, $\varphi$, or the turbulence velocity is larger than expected in 
the maser emitting regions.
The strong variation in the emission found between the observations in B and C 
configuration is probably caused by a change in the intensity of the exciting 
radiation, produced by the time evolution of the excitation of C$_2$H$_2$ and 
HCN.

A set of maser emitting arcs is 
a simple structure to explain the observations, 
but the actual morphology is probably more complex. 
A continuous
and inhomogeneous emitting region with a size larger than the synthesised
PSF or a set of emitting clumps significantly smaller than the PSF could
also produce the observed brightness distribution. 

The comparison of our observation of the vibrationally excited line 
SiS($v=1,J=14-13$) (see Figure~\ref{fig:f1} and Table~\ref{tab:table2}) with 
that by \citet{agundez_2012}, acquired with the 30-m telescope, indicates that
the peak flux of our observation in B configuration ($\simeq 2.25$~Jy with a 
HPBW~$\simeq 0.25$~arcsec; Table~\ref{tab:table2}) is $\simeq 20-30$~per~cent 
larger (\citealt{agundez_2012} reported $\simeq 1.75$~Jy with a 
HPBW~$\simeq 9.7$~arcsec).
This is not possible for a steady state line and it does not happen for the rest
of the lines observed in the current work, which are weaker than the single-dish
observations carried out by \citet{agundez_2012} and \citet{he_2008}.
Thus, line SiS($v=1,J=14-13$) might change with time as other SiS lines 
observed towards \irc{} \citep{carlstrom_1990}.
In addition, as the peak of the synthetic emission of line SiS($v=1,J=14-13$)
calculated assuming thermal emission is significantly smaller than 
the observed ($\simeq 60$~per~cent; see Section~\ref{sec:modelling.sis}), this 
line could also display maser emission, as was previously suggested by 
\citet{turner_1987} and \citet{he_2008}.

\subsection{H$^{13}$CN}
\label{sec:discussion.h13cn}

\begin{table}
\caption{Abundances of H$^{13}$CN}
\label{tab:abundances.h13cn}
\begin{tabular}{c@{\hspace{4.5ex}}c@{\hspace{4.5ex}}c@{\hspace{4.5ex}}c@{\hspace{4.5ex}}c}
\hline
\multicolumn{4}{c}{$r~(\rstar)$}    & Reference\\
$1.0-1.5$ & $1.5-5.0$ & $5-20$    & $20-50$   & \\
\hline
$8$       & $8$       & $8.0-3.3$ & $3.3-2.5$ & 1\\
$13$      & $3$       & $1$       & $1$       & 2\\
$3$       & $3$       & $11$      & $5$       & 3\\
$4$       & $4$       & $4$       & $4$       & 4\\
\hline
\end{tabular}
\medskip
\newline
All the abundances are multiplied by $10^{-7}$.
(1) This work 
(2) \citet{cernicharo_2011} 
(3) \citet{fonfria_2008} 
(4) \citet{schoier_2007}.
The data assigned to \citet{cernicharo_2011} have been calculated from their 
results for HCN assuming an isotopic ratio 
${}^{12}\textnormal{C}/{}^{13}\textnormal{C}\simeq 45$ 
\citep{cernicharo_2000,kahane_2000}.
\end{table}

The abundance of H$^{13}$CN derived from our fits agrees with the results of 
previous works throughout the envelope (Table~\ref{tab:abundances.h13cn}).
The difference between our abundance and that by \citet{fonfria_2008} is 
consequence of the lower angular resolution of their observations and the 
different distance to the star adopted in their model (180~pc) compared with 
our choice (123~pc).

\begin{figure}
\includegraphics[width=0.475\textwidth]{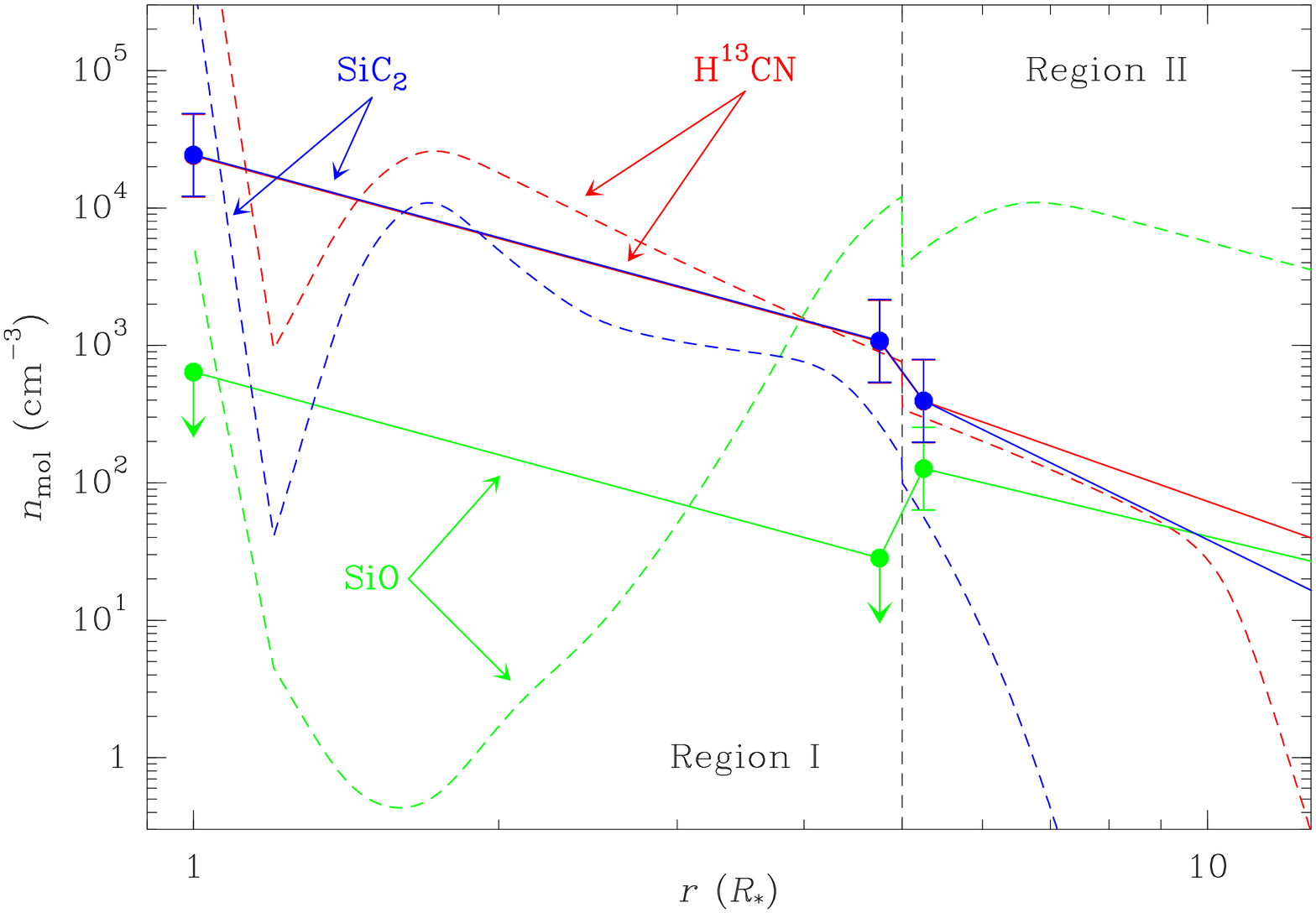}
\caption{H$^{13}$CN, SiO, and SiC$_2$ densities in Regions I and II.
The results of our fits to the observed brightness distributions
(Table~\ref{tab:table6}) are plotted in solid lines (H$^{13}$CN 
in red, SiO in green, and SiC$_2$ in blue).
The dashed curves are the results of the calculations we have performed with
the chemical model developed by \citet{tejero_1991}, which assumes TE and 
uses the most recent estimate of the H$_2$ density and the kinetic temperature 
in the dust formation zone of the envelope \citep{cernicharo_2013}.
The abundances of H$^{13}$CN and SiC$_2$ derived from our observations
are very similar in Region I and they are superimposed.}
\label{fig:f10}
\end{figure}

In Figure~\ref{fig:f10}, we compare the density of H$^{13}$CN derived from our 
fits with the results of the calculations under TE we have performed with the 
chemical model developed by \citet{tejero_1991}.
In these calculations, we have adopted the most accurate estimate of the H$_2$ 
density to date \citep{cernicharo_2013}.
Both H$^{13}$CN density profiles are compatible within our error intervals from 
$\simeq 3$ up to $10~\rstar$, where the TE condition probably fades out.
However, at the stellar surface, there is more than one order of
magnitude difference between the theoretical density provided by the
chemical model and the density obtained from the observations.
Despite the emission from Region I is unresolved in our observations,
a variation of an order of magnitude in the abundance at the stellar surface
compared with the average value in Region I would be noticeable, as the results
of our code suggest.
This situation is also found for SiO and SiC$_2$, suggesting that the H$_2$ 
density very close to the star derived by \citet{cernicharo_2013} might 
be overestimated.

We derive an average vibrational temperature in Region I 
of 2300~K by fitting the 
emission of line H$^{13}$CN($\nu_2^{1e},J=3-2$). 
This result agrees with the upper limit of the vibrational temperature
$\simeq 950-2300$~K, derived by \citet{fonfria_2008}, and is significantly 
larger compared with what \citet{cernicharo_2011} suggested ($\simeq 400$~K).
However, these differences could be the consequence of the use of larger PSFs 
than ours, tracing gas mostly at the inner acceleration shell and beyond.
The vibrational temperature derived in the current work would be better 
compared with that of the higher energy vibrational states ($\simeq 2500$~K) 
observed by \citet{cernicharo_2011}, which are very probably excited in the 
surroundings of the star.

The bipolar structure in Regions I and II derived from our fits reveals
an overall asymmetry close to the star, where
the highest abundance lays along the NNE-SSW direction with a factor of
$2-3$ larger than along the perpendicular direction.
Hence, the formation of H$^{13}$CN (HCN)
would be triggered by an anisotropic mechanism, probably related to 
the matter ejection process.

The abundance of H$^{13}$CN decreases monotonically from the 
inner to the outer acceleration shell.
In Region I, significant variations in the abundance of most molecules may 
occur due to a variety of non-equilibrium processes
\citep{willacy_1998,agundez_2006,cherchneff_2006,cernicharo_2010}.
However, molecules such as H$^{13}$CN are likely chemically inactive in Region 
II \citep{cherchneff_2006}
and they are not expected to play a dominant role in the process of dust 
formation, thus maintaining a rather constant abundance.
As we have adopted in our fits a fixed H$_2$ density profile 
($\propto r^{-2}v_\subscript{exp}^{-1}$, with a constant $v_\subscript{exp}$ out of 
the acceleration shells), the decrease observed in the H$^{13}$CN abundance 
with respect to H$_2$ in Region II may indicate that the actual H$_2$ density 
profile is steeper than what we have assumed.
This could be explained if the gas expansion velocity slightly increases 
throughout Region II, as was previously proposed by \citet{keady_1988}, 
\citet{keady_1993}, and \citet{boyle_1994}.

As it was stated above, other unknown degenerate solutions to the
fitting problem might reproduce the observations.
The parameters are more affected by uncertainties in Region I, as long
as its size is comparable to the PSF. 
However, these errors are not expected to severely change the overall structure
of the abundance distribution in Region I.
In Regions II and III, the uncertainties of the abundance are
smaller and the model is more reliable.
Globally, we expect a
good qualitative agreement between the model and the actual abundance
distribution, not only for H$^{13}$CN but also for SiO and SiC$_2$.

\subsection{SiO}
\label{sec:discussion.sio}

\begin{table}
\caption{Abundances of SiO}
\label{tab:abundances.sio}
\begin{tabular}{c@{\hspace{9ex}}c@{\hspace{9ex}}c@{\hspace{9ex}}c}
\hline
\multicolumn{3}{c}{$r~(\rstar)$} & Reference\\
$1.0-5.0$      & $5-20$    & $20-50$       & \\
\hline
$\lesssim 0.2$ & $2.4-3.3$ & $3.3-1.8$     & 1\\
$1.8$          & $1.8$     & $1.8$         & 2\\
$0.2-3.0$      & $0.2-3.0$ & $1.0$         & 3\\
$2.8$          & $2.8$     & $2.8$         & 4\\
$0.3-15$       & $15-1.7$  & $1.7$         & 5\\
$8.0$          & $8.0$     & $8.0$         & 6\\
\hline
\end{tabular}
\medskip
\newline
All the abundances are multiplied by $10^{-7}$.
(1) This work 
(2) \citet{agundez_2012} 
(3) \citet{decin_2010a}
(4) \citet{schoier_2006a}
(5) \citet{schoier_2006b} 
(6) \citet{keady_1993}.
\citet{schoier_2006b} also divided the envelope in three regions with different 
extent to ours and constant abundances.
This fact accounts for the larger variations in the abundance encountered in 
this Table in Regions I and II.
Similarly, \citet{decin_2010a} gave the abundances between the stellar surface 
and the shell at $8~\rstar$ and beyond.
\end{table}

The abundances derived from our fits agree with most of the results 
available in the literature (Table~\ref{tab:abundances.sio}).
However, the abundance proposed by \citet{keady_1993} from mid-IR observations
is a factor of $\gtrsim 3$ larger than ours throughout the envelope.
These observations were also analysed
by \citet{schoier_2006b} finding an 
abundance up to $7-8$ times larger than ours in Region II.
These significant disagreements could be the consequence of instrumental issues 
that affected these observations \citep{keady_1993}.

The comparison between our estimate of the SiO density and the calculated 
under TE suggests that this condition only stands between the surface of the
star and $\simeq 3~\rstar$ (Figure~\ref{fig:f10}).
At larger distances from the star, the SiO density under TE is up to two orders 
of magnitude larger than that derived from our observations 
suggesting the evolution of the SiO abundance is 
controlled by chemical kinetics, as was already proposed by 
\citet{agundez_2006}.

The deficit of SiO emission in front of the star noticed in the moment 0 and
the blue-shifted emission maps (about 20~per~cent of the expected emission;
Figures~\ref{fig:f3} and \ref{fig:f4}) and directly
related to a significant decrease in the SiO abundance in Region I in our 
fit (Section~\ref{sec:modelling.sio}) could be argued to be also the effect of
($i$) self-absorption or 
($ii$) absorption of the continuum emission coming from the star by SiO.
Self-absorption is produced mostly along the LoS due to absorption of the
radiation emitted by the warm gas in Region I by the colder gas in Regions 
II and III.
The different expansion velocity of the gas in Region I and Regions II and III
(5~\kms{} against 11 and 14.5~\kms, respectively; Section~\ref{sec:previous}),
allow the gas in Regions II and III to absorb less than 5~per~cent of the emission 
coming from Region I (adopting the abundance in Regions II and III from
Table~\ref{tab:abundances.sio} and an abundance for Region I equal to that 
for Region II).
Since the HPBW of the PSF of our observations in B configuration 
($\simeq 0.25$~arcsec) is more than 6 times larger than the size of the
central star of \irc{} ($\simeq 0.04$~arcsec), this
process cannot be the responsible of 
the evident emission deficit in our observation.
Regarding the second case, the stellar emission, which accounts for
the bulk of the continuum (Section~\ref{sec:discussion.continuum}), 
is absorbed by the SiO in front of the star.
This absorption is not removed from the molecular 
observations with the continuum and
could substantially modify the SiO emission in the moment 0 and blue-shifted
maps.
Our code indicates that any modification in the SiO abundance in 
Regions II and III produces a negligible emission deficit in these maps.
However, significant absorptions could arise from, e.g., dense cold clumps
in front of the star located in Region I.

The strong increase of one order of magnitude observed in the abundance between 
Regions I and II occurs around the inner acceleration zone, where dust grains 
are formed and grow.
The chemical reaction usually invoked to explain the formation of SiO in 
gas-phase in the innermost envelope of C-rich AGB stars is Si+CO$\to$SiO+C, 
efficient at high kinetic temperatures and high density 
\citep{willacy_1998,agundez_2006,cherchneff_2006,agundez_2012}.
Under these conditions, the chemical models predict an increase of $1-2$ orders 
of magnitude in the SiO abundance from the stellar photosphere up to roughly 
the inner acceleration shell, in very good agreement with our results.

The possible acceleration undergone throughout Region II
by the gas inferred from the analysis of 
the H$^{13}$CN abundance would also affect the abundance distribution derived 
from the SiO emission.
The small difference between the abundance at the inner and outer acceleration 
shells resulting from our fits (Table~\ref{tab:abundances.sio}), performed 
assuming a constant gas expansion velocity in Region II, could be an effect of 
the uncertainties of the fitting procedure.
However, the existence of a gas acceleration in Region II would enhance this 
abundance difference suggesting that SiO may be formed also throughout this
Region but at a lower rate than in the inner acceleration zone.

The SiO vibrational temperature averaged on the polar angle derived from our 
fits (1750 and 580~K in the inner and outer acceleration shells) is larger than
the kinetic temperature \citep*[$\simeq 950$ and 450~K;][]{agundez_2012}.
Thus, SiO is vibrationally out of LTE throughout the dust formation zone, 
mostly around the inner acceleration shell, where the bulk of SiO is formed.
The increment in the vibrational temperature found roughly along NE suggests 
the existence of an excitation mechanism which is not working in the rest of 
the envelope.
This mechanism may involve 
($i$) a selective excitation of the observed SiO line, or
($ii$) shocks or a strong IR or UV radiation field, which would affect the
rest of the molecules as well.
New observations of the H$^{13}$CN, SiS, and SiC$_2$ vibrationally excited 
lines at $\simeq 0.7$~arcsec resolution may help to solve this dichotomy. 

\subsection{SiC$_2$}
\label{sec:discussion.sic2}

\begin{table}
\caption{Abundances of SiC$_2$}
\label{tab:abundances.sic2}
\begin{tabular}{c@{\hspace{9ex}}c@{\hspace{9ex}}c@{\hspace{9ex}}c}
\hline
\multicolumn{3}{c}{$r~(\rstar)$}       & Reference\\
$1.0-5.0$      & $5-20$         & $20-50$        & \\
\hline
$8$            & $8.0-0.8$      & $0.8-3.4$      & 1\\
---            & ---            & $2.0-2.9$      & 2\\
$5$            & $5$            & $5$            & 3\\
$\lesssim 0.5$ & $\lesssim 0.5$ & $\lesssim 0.5$ & 4\\
\hline
\end{tabular}
\medskip
\newline
All the abundances are multiplied by $10^{-7}$.
(1) This work 
(2) \citet{cernicharo_2010} 
(3) \citet{lucas_1995} 
(4) \citet{gensheimer_1995}.
\citet{cernicharo_2010} derived the abundance of SiC$_2$ only at distances from 
the star larger than 0.5~arcsec, i.e., $25~\rstar$ for us.
\end{table}

The SiC$_2$ abundance averaged on the axial and polar angles, 
$\langle x\rangle_{\varphi\theta}$, decreases 
by an order of magnitude from the inner
to the outer acceleration zone, and then increases by a factor of 4 
throughout Region III (Tables~\ref{tab:table6}).
The decrease we observe across Region II is probably produced by
depletion of SiC$_2$ on to dust grains due to the refractory character of this 
Si-bearing molecule.
This depletion and the subsequent growth of the dust grains could enlarge their 
cross-section enough to significantly increase the force produced by the 
stellar radiation pressure on them, accelerating the dust grains.
The dust grains, dynamically coupled to the gas in Region II due to its still
high density, would also accelerate the gas, as we have suggested from the 
analysis of the abundance profile of H$^{13}$CN 
(Section~\ref{sec:discussion.h13cn}).

Our estimate of the average 
SiC$_2$ abundance in Region I agrees with most of the 
results derived from previous observations (Table~\ref{tab:abundances.sic2}), 
with the abundance estimated with chemical models under TE 
\citep*[$3-5\times 10^{-7}$;][]{willacy_1998,cernicharo_2010}, 
and with that calculated by us also under TE between $\simeq 1.5$ and 
$4~\rstar$ (Figure~\ref{fig:f10}).
Our fits suggest that
the SiC$_2$ abundance in Region I and II is larger due S than due N by
a factor of $5-6$.
The SiC$_2$ peak abundance is located to the S of the star, as in the case
of H$^{13}$CN.
This fact support the idea of an anisotropic mechanism driving the chemistry
close to the star.

The increase experienced by the SiC$_2$ abundance in Region III might be 
explained by 
($i$) the activation of a gas-phase chemical reaction or
($ii$) evaporation of SiC$_2$ molecules adsorbed to the dust grains across 
Regions I and II due to desorption processes such as grain-grain or 
molecule-grain collisions, or photodesorption triggered by the action of 
Galactic UV radiation penetrating across a clumpy envelope 
\citep{agundez_2010,decin_2010b}.
Moreover, the SiC$_2$ depletion process on to the dust grains suggested to be 
working throughout Region II would become less efficient at larger 
distances to the star as the gas density decreases and the expansion velocity 
of the dust grains grows compared to that of the gas.
Since the gas depletion is a dynamical process, inefficient desorption 
mechanisms in Regions I and II could play a major role in Region III.

\subsection{Remarkable directions in the envelope}
\label{sec:discussion.remarkable}

\begin{figure}
\centering
\includegraphics[width=0.415\textwidth]{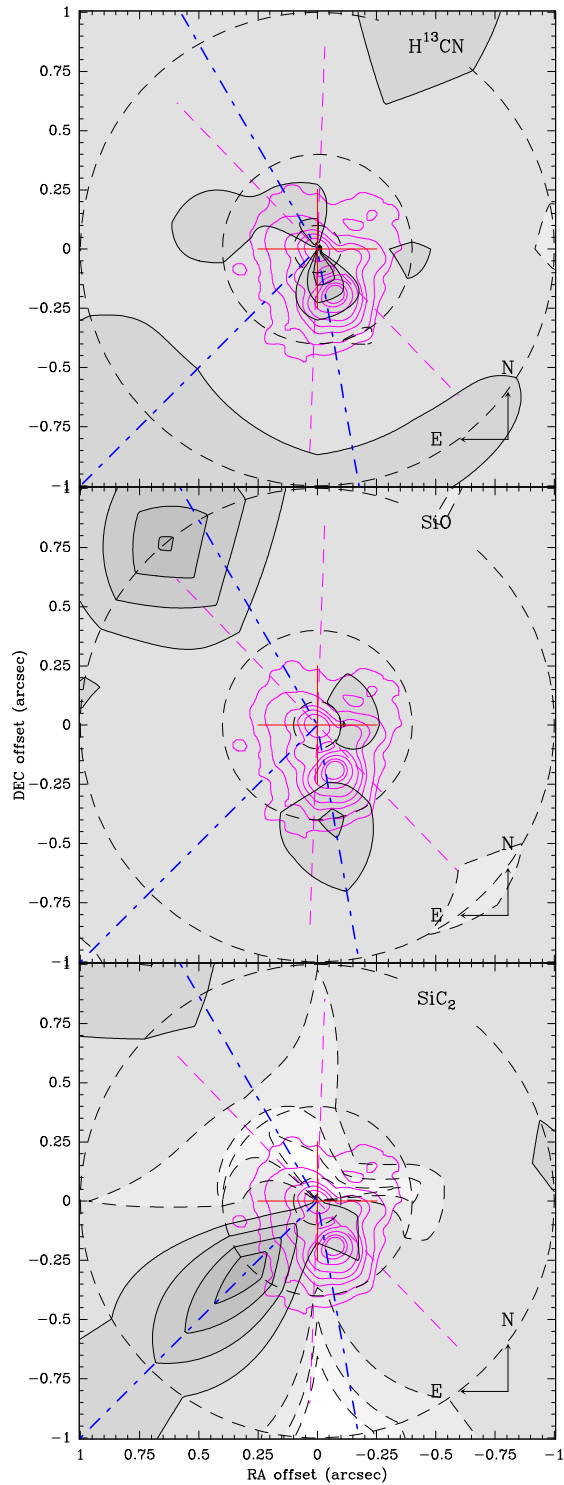}
\caption{Ratios of the abundance distributions to those averaged on the axial
and polar angles for H$^{13}$CN, SiO, and SiC$_2$ 
(grey scale and black contours).
The dash-dot straight lines (blue) represent remarkable directions (NE, S-SW, 
and SE).
The average distributions have been calculated after excluding the abundances 
along these directions.
The dashed circles are at 5, 20, and $50~\rstar$.
The solid contours are at levels 1.5, 2.0, 2.5, and 3.0.
Dashed contours are at levels 0.4, 0.5, and 0.7.
The contours plotted in magenta represent the continuum emission at $2.17~\mu$m
acquired with speckle-masking interferometry by \citet{weigelt_1998}
(Fig.~1a in this reference).
The magenta dashed lines define the cavities in the dusty component derived
by \citet{menshchikov_2001,menshchikov_2002}.}
\label{fig:f11}
\end{figure}

Figure~\ref{fig:f11} shows the relative abundance distribution of H$^{13}$CN, 
SiO, and SiC$_2$.
These maps have been obtained by dividing the abundance distributions by the 
abundance averaged on the axial and polar angles.
The relative abundance distributions of all the molecules show strong features
along directions NE and S-SW.
The H$^{13}$CN and SiO relative abundances display an excess towards these two 
directions, while the SiC$_2$ one shows a deficit.

The NE and S-SW directions are compatible with the NE-SW direction of the axis 
found in the dusty component of the envelope
\citep{dyck_1987,ridgway_1988,kastner_1994,sloan_1995,skinner_1998,haniff_1998,
weigelt_1998,weigelt_2002,osterbart_2000,tuthill_2000,tuthill_2005,leao_2006}.
From observations covering the dust formation zone, 
\citet{menshchikov_2001,menshchikov_2002} 
derived the existence of two cavities along this axis with an overall size 
$\simeq 0.5$~arcsec.
The opening angle and inclination of the cavities are $\simeq 40\degr$ and 
$\simeq 30-40\degr$, respectively, suggesting that the cavity to the SW is in 
front of the star and that to the NE is behind the star.
The axial size of all the strong features present in our relative abundance 
maps of H$^{13}$CN, SiO, and SiC$_2$ is $\gtrsim 30-40\degr$, 
comparable or larger than the 
opening angle of the cavities in the dusty component of the CSE.
Most of the molecular features exist in front of and behind the star at the 
same time with the exception of the feature of SiO at $50~\rstar$ due NE, 
which is located only behind the star.
These facts suggest that the structures associated to the molecular features 
may be wrapping the cavities in the dusty component of the envelope.

Another remarkable direction in the envelope
is SE, defined by a strong feature in the SiC$_2$ 
abundance distribution.
However, this direction seems to be unimportant for H$^{13}$CN and SiO.
It has been associated to the region throughout the CSE 
with the largest density of dust grains 
\citep*[e.g.,][]{dyck_1987,ridgway_1988,menshchikov_2001,shinnaga_2009}.
This fact joined to the agreement between the location of the cavities in the 
dusty component of the envelope along the NE-SW direction and the deficit in the
SiC$_2$ abundance might indicate that there exists a correlation between the 
density of dust grains and the SiC$_2$ abundance.

\section{Conclusions}
\label{sec:conclusions}

\irc{} has been observed at 1.2~mm with the CARMA interferometer in B and C 
configurations.
The angular resolution is one of the largest ever achieved regarding molecular
observations towards this source (HPBW~$\gtrsim 0.25$~arcsec).
The available band width allowed us to observe the continuum emission and
several lines of SiS, H$^{13}$CN, SiO, and SiC$_2$ with enough SNR to perform a 
reliable analysis.
The main conclusions of this work are:
\begin{itemize}
\item The brightness distribution of line SiS($v=0,J=14-13$) was reproduced
assuming that most of the emission comes from several maser emitting arcs.
The SiS flux measured from our two separated observations changed dramatically, 
supporting the idea of a maser origin for the SiS emission.
\item The abundance of H$^{13}$CN with respect to H$_2$ shows a bipolar
structure along the NNE-SSW direction in Regions I and II.
In average, it undergoes a significant decrease
between the inner and outer acceleration shells.
The vibrational temperature close to the star is about 700~K
in average larger than was previously suggested.
\item The SiO appears mostly in the inner acceleration shell.
Inwards, its abundance is at least
one order of magnitude smaller than for the rest of the envelope.
Most of the SiO is located behind the star.
The vibrational temperature indicates that SiO is vibrationally out of LTE
between the inner and outer acceleration shells.
A substantial increment of this magnitude has been found around the NE compared
to other directions.
\item The abundance of SiC$_2$ shows a significant minimum located at the outer 
acceleration zone, probably produced by depletion on to dust grains.
A further increase in the abundance has been detected beyond the outer 
acceleration shell.
Most of the SiC$_2$ is located towards the S of the star in Regions
I and II.
\item The abundance distributions of H$^{13}$CN, SiO, and SiC$_2$ are irregular 
and show remarkable directions along which the abundances are significantly 
different from the rest of the envelope.
These directions are NE, S-SW, and SE.
The directions NE and S-SW match up with the axis of the dusty component of 
the envelope (NE-SW), previously inferred from continuum observations.
The direction SE coincides with that associated to the region of the envelope
with the largest amount of dust.
The abundance of SiC$_2$ seems to be correlated with the density of dust grains.
\end{itemize}

\appendix

\section{Details of the numerical code}
\label{sec:appendix}

The code used in the current work is an improved 3D version of the
1D code developed by \citet{fonfria_2008} capable of reproducing
the molecular and continuum emissions that
come from symmetric or asymmetric circumstellar envelopes composed of 
expanding gas and dust.
Our code solves the radiation transfer equation along the LoS for a 
number of positions in the plane of the sky 
and produces synthetic line emission data cubes adopting a given physical
and chemical model for the envelope, i.e., the H$_2$ density, 
molecular abundance, gas expansion velocity, line width,
and rotational and vibrational temperature spatial distributions.
The procedure followed to perform the calculations 
\textit{does not} rely on the LVG approximation and 
the code is well adapted to 
model the emission of the inner layers of the envelope,
where the gas density and the temperatures are
high, and the expansion velocity gradient
is supposed to be small.
In this Section we benchmark this new version of the code.

\subsection{Gridding and sampling of the physical and chemical 
magnitudes}
\label{sec:appendix.gridding}

The system composed of the central star and the circumstellar envelope 
is described by spherical coordinates.
The envelope is divided into concentric shells
centred on the star in order to sample the magnitudes describing the physical
and chemical conditions.
To solve the radiation transfer equation the envelope is discretised
into a set of right cylinders with their axes matching up with the LoS
passing through the central star.
The combination of both structures gives rise to a set of 
annular regions parallel to the plane of the sky.
A number of \textit{principal} axial or position angles, $\varphi$, are
selected to define the same number of sets of sections of 
the annular regions or \textit{cells}.
In each of these sets of cells, any physical or chemical magnitude is sampled
depending on the distance to the star, $r$, and the polar angle, $\theta$.
The length of each cell along the LoS is controlled by the number of shells. 
The radiation transfer equation is solved along the LoS 
for every principal axial angle and impact parameter.
The emission for any other axial angle is calculated by linear interpolation.
The number of shells and principal axial angles are chosen to produce
maps independent on the sampling within an error of 10~per~cent of the 
observational uncertainty.
About 100 shells and $9-12$ principal axial angles have been required to
reproduce the observations presented in the current paper.

\subsection{A brief description of the program}
\label{sec:appendix.program}

Contrarily to many other codes developed to reproduce the emission
of circumstellar envelopes or molecular clouds
\citep*[e.g.,][]{gonzalezalfonso_1993,dullemond_2000,vanzadelhoff_2002,vandertak_2007,daniel_2008}, 
\textit{our code does not solve the statistical equilibrium equations} (SE).
The number of levels involved in the calculations 
regarding the warmest regions of the envelopes of AGB stars 
($T_k\simeq 1000-3000$~K)
ranges between several hundred for the lightest molecules 
(e.g., CO, SiS, SiO, CS)
to several thousand for more complex species (e.g., H$_2$O, HCN, C$_2$H$_2$,
C$_2$H$_4$).
Resolving the SE for such an environment is a very challenging task due to
the lack of collisional coefficients and the huge computing effort demanded.
Hence, we compute the populations of the molecular levels
assuming a Boltzmann distribution
with rotational and vibrational temperatures depending on the
ro-vibrational quantum numbers.
The partition function, necessary to calculate the opacity of the 
lines to be modelled, 
is computed by direct summation over all the considered levels.
Its accuracy depends on the molecule and the
uncertainty is usually $1-5$~per~cent for diatomic,
linear, and symmetric molecules, and up to $10-15$~per~cent
for asymmetric molecules
as a consequence of the complexity of the calculations and 
the lack of spectroscopic constants in the literature.

Once the populations of all the required levels are calculated throughout
the whole envelope,
the radiation transfer equation is solved along the LoS
assuming that the 
physical and chemical magnitudes in each cell of the grid are constant.
Hence, the intensity emerging from
a given point in the plane of the sky is
\begin{equation}
I_\nu = B_\nu(T_\subscript{CMB})e^{-\sum_{j=1}^n \tau_{\nu,j}}
 +\sum_{i=1}^n S_{\nu,i}\left(1-e^{-\tau_{\nu,i}}\right)e^{-\sum_{j=1}^{i-1} \tau_{\nu,j}},
\end{equation}
where we have considered the CMB continuum and
$n$ is the number of cells along the LoS at that point in the plane
of the sky.
The source function, $S_\nu$, and the optical depth, $\tau_\nu$, are evaluated
in every cell, where the first and $n$-th cells and the
closest and farthest cells to Earth, respectively.
The optical depth and the source function are defined as
\begin{equation}
\tau_\nu=\sum_{i=1}^{m} \tau_{\nu,i}
\end{equation}
and
\begin{equation}
S_\nu=\sum_{i=1}^{m}\frac{k_{\nu,i}}{\sum_{j=1}^m k_{\nu,j}} B_{\nu,i},
\end{equation}
where $m$ is the number of lines of a single molecule to be reproduced and
$k_\nu$ is the line opacity.
All the magnitudes in these equations depend on the rotational and vibrational
temperatures, and on the gas expansion velocity.
The addition of more molecules or dust is straighforward.
This approach allows for a correct modelling of blended lines
regardless of any velocity gradient or the
effect of dust on the molecular emission, important in the mid-infrared range.
Moreover, our code is able to compute thermal or maser emission 
from any point of the envelope just by chosing the correct rotational
temperatures, that can be negative.

\subsection{Benchmarking}
\label{sec:appendix.benchmark}

The performance of our code
has been analysed assuming different symmetries (1D, 2D, and
3D) by comparing the results with those of ad-hoc codes for
simple scenarios
or with those obtained
from a non-local, non-LTE code \citep{daniel_2008}.
The uncertainties for the 1D and 2D problems
are defined for all the comparisons below as
$(F_\subscript{ours}-F_\subscript{control})/\max{(F_\subscript{control})}$,
where $F_X$ is the emitted flux, $X$ is the code (ours or the
control codes ad-hoc or
DC, i.e., \citealt{daniel_2008}), and $\max{(F_\subscript{control})}$ is the
maximum value of the flux calculated with the control code.

\subsubsection{1D problem}
\label{sec:1dproblem}

\begin{figure}
\centering
\includegraphics[width=0.475\textwidth]{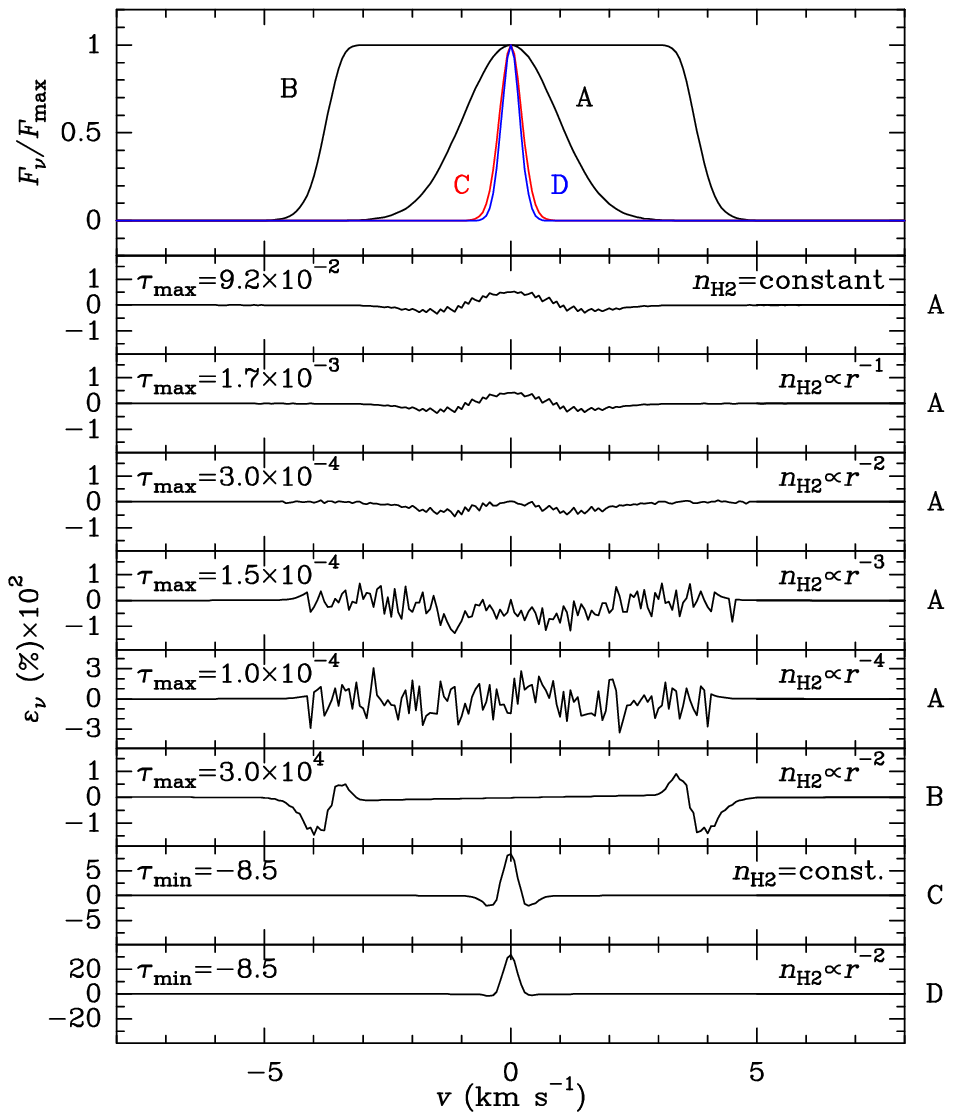}
\caption{Test line normalised emission from a static isothermal 
spherically symmetric cloud depending of the H$_2$ density profile.
The upper insert contains the emission 
for the optically thin (A) and optically thick (B) thermal
emission and for the maser emission (C, in red, and D, in blue).
The line profile in the thermal
optically thin case is independent of the H$_2$ density.
The maximum flux for the thermal lines ranges between 
$3\times 10^{-6}$ and 2.5~Jy, and between $5\times 10^7$ and $10^8$~Jy 
for the maser lines.
All the spectra have been calculated with an ad-hoc code developed for this 
purpose.
The optical depth
for every impact parameter has been analytically calculated.
The continuum due to the CMB has been removed.
The residuals, $\varepsilon_\nu$, are defined as 
$(F_\subscript{ours}-F_\subscript{ad-hoc})/\max(F_\subscript{ad-hoc})$, where
$F_\subscript{ad-hoc}$ and $F_\subscript{ours}$ are the synthetic fluxes 
calculated with the ad-hoc code and ours, respetively.}
\label{fig:f12}
\end{figure}

\begin{figure}
\centering
\includegraphics[width=0.475\textwidth]{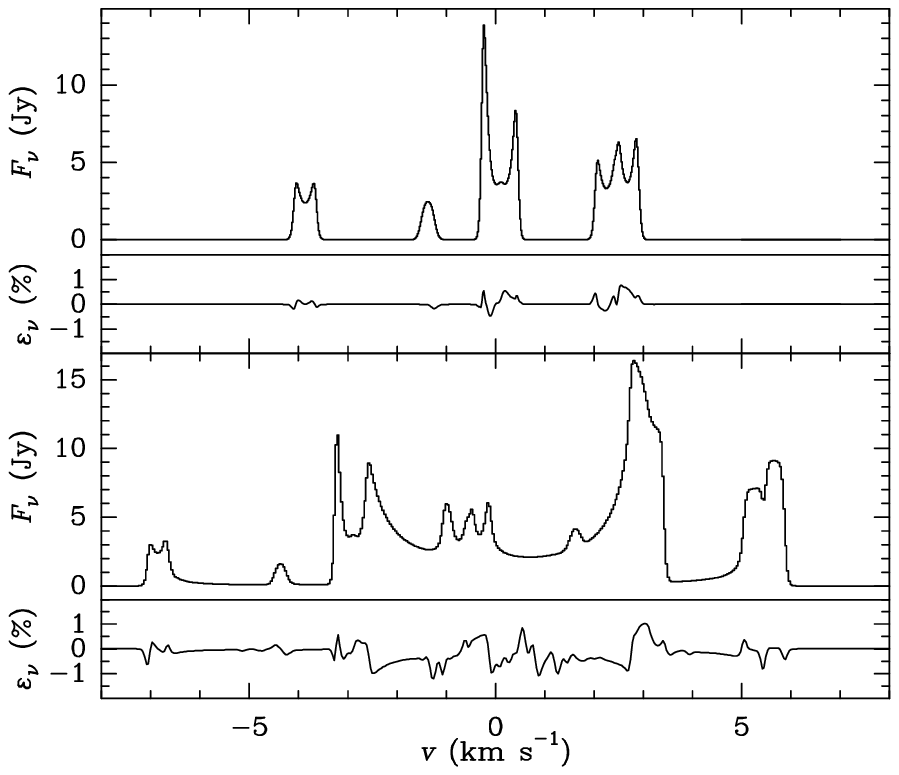}
\caption{Hyperfine structure of line H$^{13}$CN($v=0,J=2-1$) towards a
spherically symmetric cloud expanding at 0 and 3~\kms{}
(upper and lower inserts, respectively).
See the text for a description of the physical and chemical properties of
the cloud.
The plotted spectra have been calculated with the non-local, non-LTE
code developed by \citet{daniel_2008}.
The continuum due to the CMB has been removed.
The residuals, $\varepsilon_\nu$, are defined as 
$(F_\subscript{ours}-F_\subscript{DC})/\max(F_\subscript{DC})$, where
$F_\subscript{DC}$ and $F_\subscript{ours}$ are the synthetic fluxes 
calculated with the code by \citet{daniel_2008} and ours, respetively.}
\label{fig:f13}
\end{figure}

The abundance and temperatures are allowed to vary with
the distance to the star, $r$.
Three scenarios involving thermal and maser emission 
have been used to compare the results of our code with
partially analytical solutions to simple problems (Scenario 1
for thermal emission and Scenario 2 for maser emission) and with 
totally numerical results of the non-local
non-LTE code developed by \citet{daniel_2008} (Scenario 3; thermal emission).
\begin{description}
\item[\textit{Scenario 1:}] 
We have calculated the emission of a test line ($A_{ul}=10^{-8}$~s$^{-1}$,
$B=3$~cm$^{-1}$, $J=1-0$) from
an isothermal spherically symmetric envelope assuming LTE, $T_k=100$~K,
an H$_2$ density profile following the power-law $r^{-\alpha}$, 
where $\alpha=0,\ldots,4$, and thermal line width ($\simeq 2.14$~\kms;
Figure~\ref{fig:f12}, lines A and B).
The maximum optical depth ranged from $10^{-4}$ to $0.1$.
An extremely optically thick case 
($\tau_\subscript{max}\simeq 3\times 10^4$) 
has been also considered with a H$_2$ density profile $\propto r^{-2}$.
The optical depth for every impact parameter has been analytically calculated.
The convolution of the emission 
with a HPBW~$=1$~arcsec has been performed numerically with the 
trapezoidal rule.
The code accurately computes the emission from clouds with significantly
different H$_2$ density profiles independently of the optical depth,
even in the extremely optically thick regime.
The results of our code deviate from the control solutions 
in less than 0.01~per~cent.
\item[\textit{Scenario 2:}] 
We have calculated the emission of the test line from the
same cloud than in Scenario 1 but assuming
an excitation temperature of $-100$~K for the modelled line
throughout the whole cloud (Figure~\ref{fig:f12}, lines C and D).
The minimum optical depth was $-8.5$.
The optical depth has been computed analytically.
In spite of the large emission of the line, with a maximum ranging between
$5\times 10^7$ and $10^8$~Jy, the relative error is smaller than 0.4~per~cent.
\item[\textit{Scenario 3:}] 
We have calculated the emission of the hyperfine 
structure of lines H$^{13}$CN($v=0,J=1-0$) and H$^{13}$CN($v=0,J=2-1$) coming
from an isothermal spherically symmetric envelope with $T_k=10$~K and
a constant H$_2$ density (Figure~\ref{fig:f13}).
H$^{13}$CN is out of LTE throughout the whole envelope, mostly 
in the outer shells.
A turbulence velocity of 0.1~\kms{} and the thermal linewidth have been
considered resulting in a total line width of $\simeq 0.2$~\kms.
The expansion velocity has been assumed to be 0, 3, and 10~\kms.
The maximum 
optical depth of the components of the hyperfine structure ranges betwees
$\simeq 1$ and 60.
Our code is capable of dealing at the same time
with blended lines with quite different optical depths and properly 
modelling self-absorption.
The results of our code differs from those of the control code in less than
1~per~cent for all the components of the hyperfine structure of the modelled
lines, regardless of the gas expansion velocity.
\end{description}

The good agreement between the results of our code and those of 
the control codes
indicates that the populations of the levels involved in the calculations
are well determined under LTE or out of LTE, and 
the methodology followed to resolve the radiation transfer
equation is properly implemented and works for a variety of situations,
including the modelling of emission from systems
displaying maser emission.
Thus, our code is expected to give accurate results also in more complex 1D 
scenarios.

\subsubsection{2D problem}
\label{sec:appendix.2dproblem}

In this Section we show that our 
code is capable of dealing with abundances
and excitation temperatures depending on $r$ and
the polar angle, $\theta$.

\begin{figure}
\centering
\includegraphics[width=0.475\textwidth]{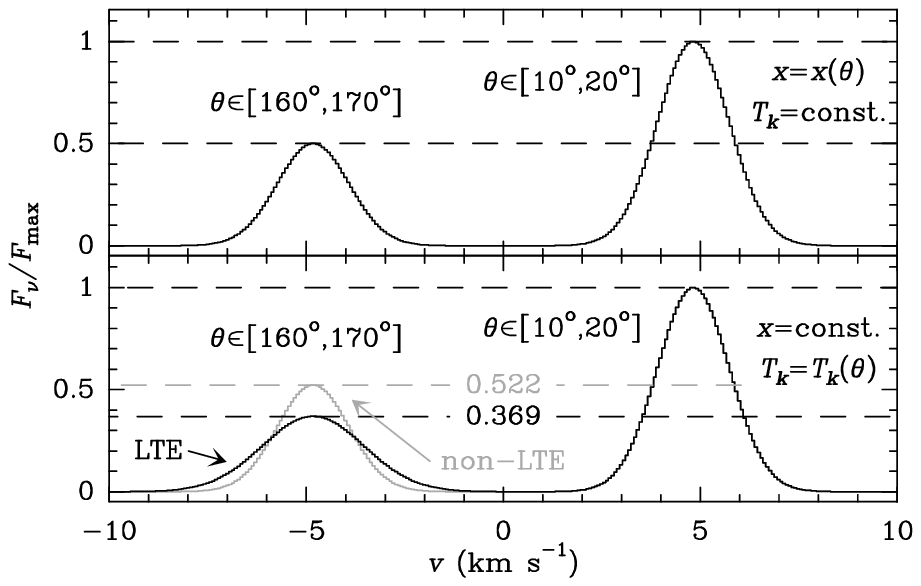}
\caption{Test line normalised to its maximum from a
cloud composed of two hollow cones
placed symmetrically behind ($10\degr\le\theta\le20\degr$,
back cone) and in front of the star ($160\degr\le\theta\le170\degr$, front 
cone).
The gas expansion velocity is set to 5~\kms{} and
the H$_2$ density profile is $\propto r^{-2}$.
The abundance is set to zero out of the cones and between 1 and 7~\rstar{}
to avoid shadowing effects from the star.
The cloud is assumed to be under LTE and it is optically thin.
The spectrum in the upper insert is calculated with an abundance of
$10^{-7}$ and $5\times 10^{-8}$ in the back and front cones, respectively,
and a constant $T_k=100$~K (LTE).
The contribution of the front cone is half of the back cone.
The black spectrum in the lower insert results from assuming an abundance of 
$10^{-7}$ in both cones and $T_k=100$ and 200~K in the back and front cones,
both under LTE.
The grey spectrum is calculated under LTE in the back cone with
$T_k=100$~K and out of LTE
with $T_k=100$~K and $T_\subscript{exc}=200$~K in the front cone.
Theoretically, the ratio of the integrals of the contributions of the
front to the back cones should be of $\simeq 0.522$ for both cases.
The ratios of the peaks of the contributions are $\simeq 0.369$ and 0.522
for the LTE and non-LTE scenarios, respectively.
The differences between the theoretical values stated above and the 
results calculated with
our code are smaller than 0.05~per~cent.}
\label{fig:f14}
\end{figure}

Figure~\ref{fig:f14} shows the emission of a test line 
(see Section~\ref{sec:1dproblem}) coming from an expanding envelope 
($v_\subscript{exp}=5$~\kms, $n_\subscript{H$_2$}\propto r^{-2}$,
thermal line width, optically thin regime)
with zero abundance except in
two hollow truncated right circular cones with their axes aligned with the LoS
(back and front cones) 
where the abundance is non-zero in three different situations:
\begin{description}
\item[\textit{Scenario 1:}] The abundance is twice smaller in the 
front cone than
in the back cone, $T_k=100$~K in both, and the envelope is under LTE.
The contribution from the front cone is the same than from the back cone but
scaled by a factor of 0.5, as expected.
The discrepancies between the theory and the results of our code are 
0.04~per~cent.
\item[\textit{Scenario 2:}] The abundance of both cones is equal but 
$T_k=100$ and 200~K in the back and front cones, respectively.
The envelope is under LTE.
The ratio of the integrated flux of the front cone to that of the back cone
calculated from the results of our code is in very good agreement with the
theoretical predictions within an uncertainty of 0.05~per~cent.
\item[\textit{Scenario 3:}] The abundance of both cones is equal but the 
back cone is under LTE with $T_k=100$~K and the front cone is rotationally
out of LTE with $T_k=100$~K and 
$T_\subscript{exc}=200$~K for the modelled line.
The ratio of the integrated flux of the front cone to that of the back cone
calculated from the results of our code agrees with
theoretical predictions with an error of 0.05~per~cent.
\end{description}

\begin{figure}
\centering
\includegraphics[width=0.475\textwidth]{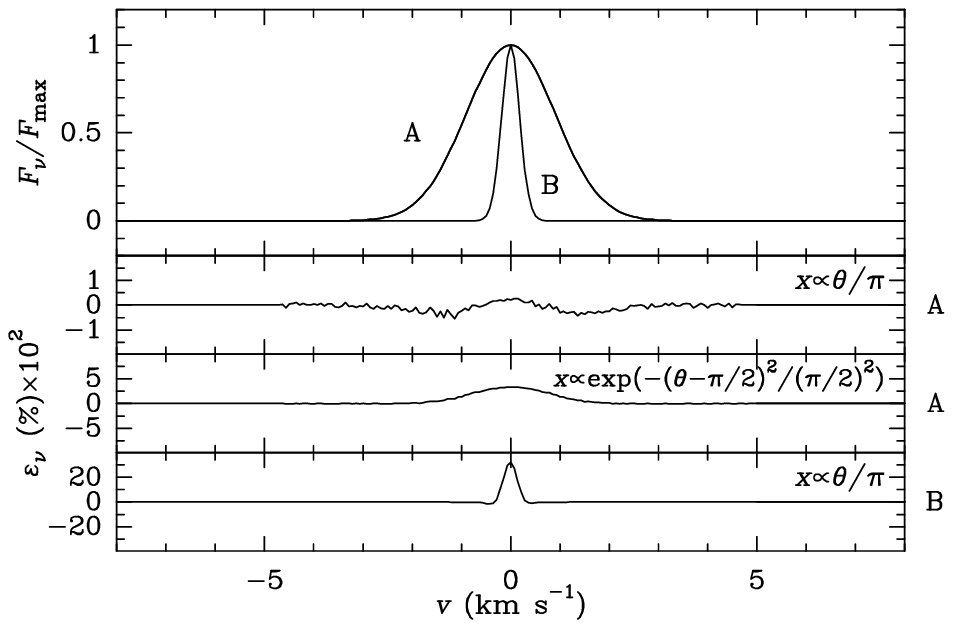}
\caption{Test line normalised to its maximum from a static isothermal cloud
with the abundance $\propto \theta/\pi$ and 
$e^{-(\theta-\pi/2)^2/(\pi/2)^2}$ (thermal emission, A, and maser
emission, B).
The H$_2$ density profile is $\propto r^{-2}$.
The thermal emission (A) is calculated under LTE with $T_k=100$~K
while the maser emission (B) is computed by assuming $T_k=100$~K and
a $T_\subscript{exc}=-100$~K for the test line.
In the thermal emission case,
the normalised line profile is the same for both abundance distribution.
The CMB continuum has been removed.
The residuals, $\varepsilon_\nu$, are defined as in Figure~\ref{fig:f12}.}
\label{fig:f15}
\end{figure}

The code can deal with more complex continuous abundance distributions 
depending on the polar angle.
In Figure~\ref{fig:f15}, we have plotted the emission of the test molecule
coming from a static isothermal envelope under LTE 
(thermal emission; $n_\subscript{H$_2$}\propto r^{-2}$, $T_k=100$~K, 
thermal line width $\simeq 2.1$~\kms)
and under LTE but with an excitation temperature for the computed line
of $-100$~K (maser emission).
The adopted abundance distributions are proportional to
$\theta/\pi$ and $e^{-(\theta-\pi/2)^2/(\pi/2)^2}$.
The emission have been compared to the results of an ad-hoc code in which
the integral of the
optical depth for each impact parameter has been calculated by using
the trapezoidal rule.
The difference between the results of both codes 
for the thermal emission is smaller than 0.05~per~cent if
$\tau_\subscript{max}\lesssim 10^4$
and smaller than 0.4~per~cent 
for the maser emission ($\tau_\subscript{max}\simeq -17$ with a
maximum flux of $9\times 10^7$~Jy).

The good results of these tests reveal that the sampling of the abundance
and the temperature distributions are well implemented regarding the 
polar angle.
The populations of the molecular levels are accurately calculated under
LTE and out of LTE and the
methodology to solve the radiation transfer equation works fine also for
an envelope depending on $r$ and $\theta$, regardless of
the optical depth of the modelled line.

\subsubsection{3D problem}

\begin{figure}
\centering
\includegraphics[width=0.4\textwidth]{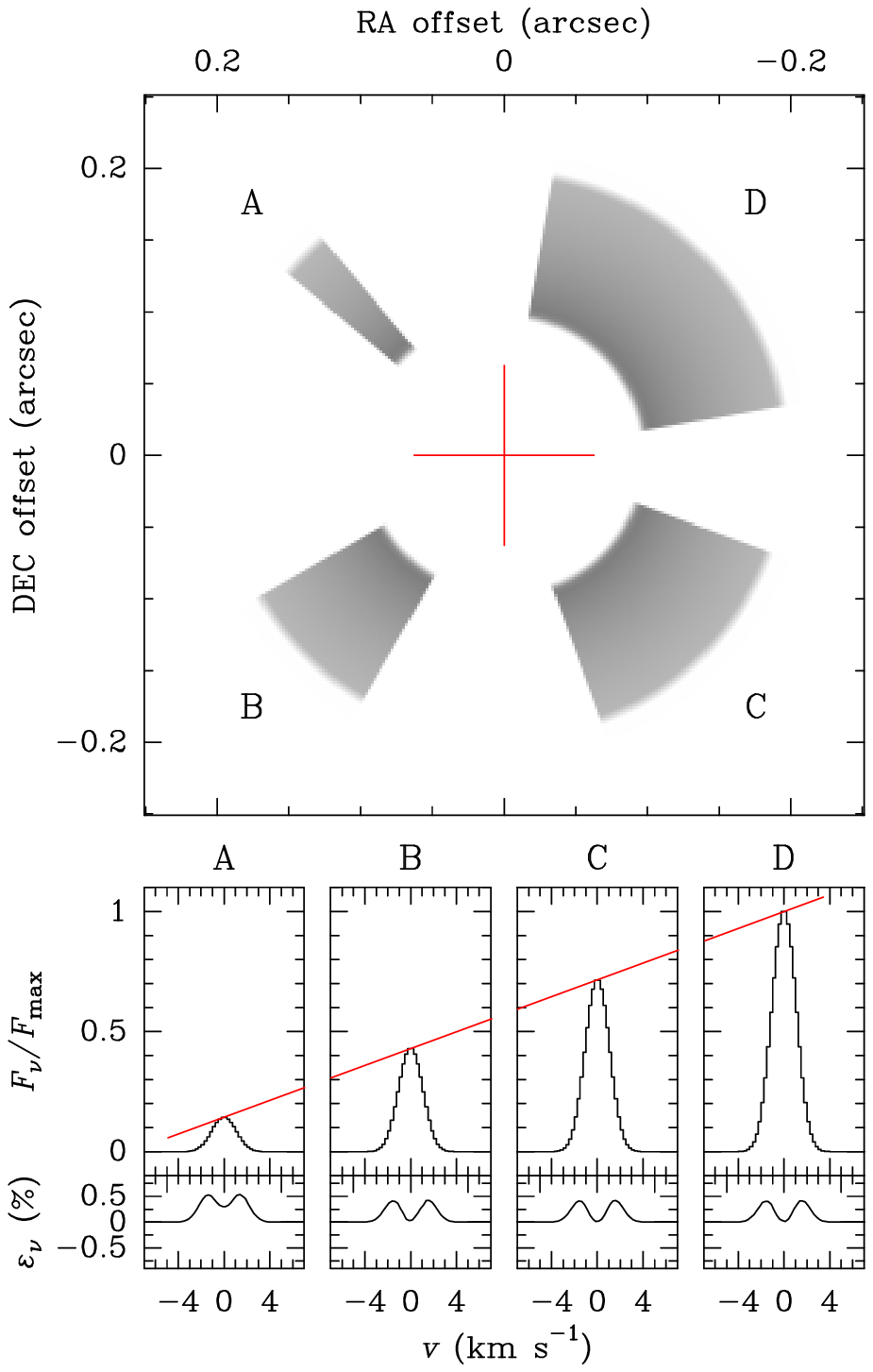}
\caption{Unconvolved 
moment 0 map of the emission of the test line showing a structure 
in the plane of the sky (upper insert).
The components of the structure, labelled A, B, C, and D,
are defined in the ranges of the axial angle
$[40\degr,50\degr]$, $[120\degr,150\degr]$, $[200\degr,250\degr]$, and
$[280\degr,350\degr]$.
The polar angle varies between 80\degr{} and 100\degr{} for the whole structure.
The distance to the star ranges from 5 to 10~\rstar.
The abundance in the rest of the envelope is zero.
The H$_2$ density profile is $\propto r^{-2}$.
The kinetic temperature is 100~K and the envelope is under LTE.
The gas expansion velocity is 5~\kms.
The emission is optically thin with $\tau\lesssim 0.4$.
The spectra of the components of the structure are plotted in the lower
inserts.
The red line shows that the ratio of the flux of the lines linearly depends,
as expected, on the axial angular size of the components.
The uncertainty, $\varepsilon_\nu$, is derived by comparing the spectrum of
each component with the scaled spectrum
produced by a complete ring with the same conditions (2D problem).}
\label{fig:f16}
\end{figure}

In the 3D problem,
the physical and chemical magnitudes are allowed to vary with the three
spherical coordinates, $(r,\varphi,\theta)$, at the same time.

In Figure~\ref{fig:f16}, a simple structure to show the performance
of the code is plotted.
The structure comprises four components (A, B, C, and D)
that revolve 10\degr{} in the polar angle around the plane of the sky.
The axial angle size depends on the component (10\degr, 30\degr, 50\degr,
and 70\degr, respectively).
The physical and chemical conditions are the same in every component.
The total flux emitted by each
component depends linearly on the axial angular size,
as expected.
This result holds for the optically thick regime 
since the power to solve
the radiation transfer equation showed in the 2D problem
(Section~\ref{sec:appendix.2dproblem}) is inherited in the 3D case.
We can compare the flux of the components 
with that of a continuous ring
with the same physical and chemical conditions than the components
treated as a 2D problem by scaling the flux of the ring with the
ratio of the sizes.
The agreement is better than 0.5~per~cent.

The ability of the code to define 3D structures 
depending on the coordinates ($r,\varphi,\theta$) and
to calculate the populations of the molecular levels and to
solve the radiation transfer equation along 2D structures with
coordinates ($r,\theta$)
makes our code suitable to accurately reproduce the 3D emission of 
any molecule regardless of the physical and chemical conditions of the emitting
cloud.

\section*{Acknowledgements}

We thank all members of CARMA staff that made observations possible. 
We also thank F. Daniel because of his invaluable 
help on the testing of the code used to fit the observed molecular emission,
G. Weigelt and collaborators for the use of Fig.~1a in \citet{weigelt_1998},
and the anonymous referee for helpful comments and suggestions.
During part of this study, J. P. F. was supported by the UNAM through a
postdoctoral fellowship. 
He thanks S. Torres-Peimbert because of her kind support.
M. F.-L. acknowledges financial support from University of Illinois and the 
hospitality of the UNAM.
C. S.-C. and J. C. have been partially  supported by the Spanish MINECO 
through grants CSD2009-00038, AYA2009-07304, and AYA2012-32032.

\end{document}